\documentclass[%
 reprint,
superscriptaddress,
amsmath,amssymb,
aps,
]{revtex4-2}

\usepackage{graphicx}
\usepackage{dcolumn}
\usepackage{bm}
\usepackage{hyperref}
\usepackage{verbatim} 
\hypersetup{
colorlinks,
citecolor=blue,
filecolor=blue, 
linkcolor=blue, 
urlcolor=blue
}

\usepackage{color}
\usepackage{xcolor}
\usepackage{ulem}

\newcommand{\yambo}{\textsc{yambo}}
\newcommand{\qe}{\textsc{QuantumEspresso}}

\renewcommand{\emph}{\textit}
\newcommand{\suppinfo}{Supplemental Material~\cite{supp-info}}
\newcommand{\ip}{\mathrm{Im}}
\newcommand{\rp}{\mathrm{Re}}

\begin{document}

\preprint{APS/123-QED}

\title{
Unraveling many-body effects in ZnO: Combined study using momentum-resolved electron energy-loss spectroscopy and first-principles calculations
}

\author{Dario A. Leon}
\email{dario.alejandro.leon.valido@nmbu.no}
\affiliation{
 Department of Mechanical Engineering and Technology Management, \\ Norwegian University of Life Sciences, NO-1432 Ås, Norway
}%

\author{Cana Elgvin}
\affiliation{Centre for Materials Science and Nanotechnology, Department of Physics, University of Oslo, NO-0316 Oslo, Norway
}%

\author{Phuong Dan Nguyen}
\affiliation{Centre for Materials Science and Nanotechnology, Department of Physics, University of Oslo, NO-0316 Oslo, Norway
}%

\author{\O ystein Prytz}
\affiliation{Centre for Materials Science and Nanotechnology, Department of Physics, University of Oslo, NO-0316 Oslo, Norway
}%

\author{Fredrik S. Hage}
\affiliation{Centre for Materials Science and Nanotechnology, Department of Physics, University of Oslo, NO-0316 Oslo, Norway
}%

\author{Kristian Berland}
\affiliation{
 Department of Mechanical Engineering and Technology Management, \\ Norwegian University of Life Sciences, NO-1432 Ås, Norway
}

\date{\today}

\begin{abstract}
We present a detailed study of the dielectric response of ZnO using a combination of low-loss momentum-resolved electron energy-loss spectroscopy (EELS) and first-principles calculations at several levels of theory, from the independent particle and the random phase approximation with different variants of density functional theory (DFT), including hybrid and DFT$+U$ schemes; to the Bethe-Salpeter equation (BSE). 
We use a method based on the $f$-sum rule to obtain the momentum-resolved experimental loss function and absorption spectra from EELS measurements.
We characterize the main features in the direct and inverse dielectric functions of ZnO and their dispersion, associating them to single-particle features in the electronic band structure, while highlighting the important role of many-body effects such as plasmons and excitons. We discuss different signatures of the high anisotropy in the response function of ZnO, including the symmetry of the excitonic wave-functions.
\end{abstract}

\maketitle

\section{Introduction}
Zinc oxide (ZnO) is a transparent semiconductor material with attractive properties for electronics, spintronics, optoelectronics, and nanotechnology~\cite{Ozgur2005JAP,Klingshirn2007pssb,Wang2007Micron,Ozgur2010PIEEE,Klingshirn2010pssb}. It has been the subject of intense research in
diverse applications, including transistors, supercapacitors, lasers, solar cells,  batteries, photoluminescent materials, photocatalysis and biosensors (see e.g. recent reviews~\cite{Theerthagiri2019Nanotech,Borysiewicz2019Cryst,Wibowo2020RSCadv,Sharma2022MTP}), 
 and beyond, being used in sunscreens and cosmetics~\cite{Ginzburg2021PPS,Raha2022NanoAdv}, textiles and rubber industry~\cite{Raha2022NanoAdv},  pharmaceutics~\cite{Carmona2018Nanomat}, and many other biomedical and biological applications~\cite{Sirelkhatim2015NML,Theerthagiri2019Nanotech,MedinaCruz2020JPM,Raha2022NanoAdv}. 
With this wide range of applications, there is a great need for fundamental understanding of its electronic and optical properties. In fact, ZnO has been characterized with a range of different spectroscopic techniques~\cite{Srikant1998JAP,Ozgur2005JAP}, including optical absorption~\cite{Thomas1960JPCS,Park1966PR,Srikant1997JAP}, transmission~\cite{Liang1968PRL}, reflection~\cite{Thomas1960JPCS,Park1966PR}, photoreflection~\cite{Chichibu2003JAP}, photoemission~\cite{Powell1971PRL,Girard1997SurfSci,Preston2008PRB,Kobayashi2009JAP,Lim2012PRB}, ellipsometry~\cite{Jellison1998PRB,RakelIBOOK2008}, photoluminescence~\cite{Mang1995SSC,Teke2004PRB,Tsoi2006PRB}, electron energy-loss spectroscopy (EELS)~\cite{Dorn1977PRB,Ding2005JEM,Zhang2006Micron,Wang2007Micron,Wu2012Plasmonics}; in combination with characterization methods such as X-ray~\cite{Recio1998PRB,Desgreniers1998PRB}, low-energy electron diffraction~\cite{Girard1997SurfSci} and other microscopy techniques~\cite{Ozgur2005JAP}. 

ZnO has also attracted considerable interest from the {\it ab-initio} community due to the difficulty of reproducing its experimental quasi-particle (QP) band structure from first principles. At the density functional theory (DFT) level, calculations with standard semi-local functionals such as the local density approximation (LDA) or the generalized gradient approximation (GGA) underestimate the electronic bandgap of ZnO at zero Kelvin by as much as 80\% (see e.g. Refs.~
\cite{Berger2012PRB,Stankovski2011PRB,Shishkin2007PRB}).
At the $GW$ level of theory, converging QP energies has proven to be very 
cumbersome, with a strong interdependency of the convergence parameters~\cite{Shih2010PRL,Friedrich2011PRB,Rangel2020CPC}. 
Although several schemes to accelerate the calculations with respect to the number of empty states have been proposed~\cite{Bruneval2008PRB,Berger2012PRB,Deslippe2013PRB,Klimes2014PRB}, $G_0W_0$ calculations on top of LDA or GGA still underestimate the bandgap of ZnO by around 10\%. This underestimation arises when evaluating the self-energy both with plasmon-pole models and with numerical full-frequency integrations~\cite{Usuda2002PRB,Stankovski2011PRB,Miglio2012EPJB,Larson2013PRB}. 

Such $G_0W_0$ methods also find too shallow $d$-band positions compared to photoemission measurements~\cite{Powell1971PRL,Lim2012PRB}, which can be attributed to a spurious hybridization of Zn-$s$/O-$p$ and Zn-$d$ states. This hybridization has also been linked to the bandgap underestimation~\cite{Usuda2002PRB,Lim2012PRB,Lany2013PRB,Gruneis2014PRL}.
While quasi-particle self-consistent $GW$ schemes~\cite{Preston2011PRB,vanSchilfgaarde2006PRL,Shishkin2007PRL,Jiang2016PRB,Kutepov2017CPC} can improve the bandgap, these schemes 
do not update the wave-functions and hence, do not correct their 
orbital hybridizations~\cite{Lim2012PRB}. 
Fully self-consistent $GW$ schemes can correct both, the position and the hybridization of the $d$-states; however, such calculations are computationally expensive and tend to overestimate bandgaps due to the lack of vertex corrections~\cite{Shishkin2007PRL,Chen2015PRB,Cao2017PRB}. 
Incorporating such vertex corrections bears prohibitively higher computational costs. 
To address these issues at a more affordable level of theory, several DFT and $GW$@DFT schemes using hybrid functionals or including a Hubbard-like $U$ term (DFT$+U$) have been applied to ZnO~\cite{Harun2020ResInPhys,Fuchs2007PRB,Gori2010PRB,Yan2011SST,Lim2012PRB,Kang2014PRB,Ciechan2021SREP,Agapito2015PRX,Gopal2015PRB,Riefer2017JPCM,Zhang2018PRB,Harun2020ResInPhys,Colonna2022JCTC,Goh2017CMS}. 
While the $U$ term can be a simple and effective way to account for the strong on-site Coulomb interactions of localized electrons like the Zn-$d$ states, there is no unique way to choose this parameter~\cite{Harun2020ResInPhys}.

Alongside the band structure, the response function of pure and doped ZnO has also been studied both experimentally~\cite{Mang1995SSC,Ding2005JEM,Wang2005APL,Zhang2006Micron,Wang2007Micron,Wang2008Micron,Wu2012Plasmonics,Yeh2021PRApp} and theoretically~\cite{Zhang2006Micron,Gori2010PRB,Zhang2018PRB,Harun2020ResInPhys,Yeh2021PRApp}. 
Single-particle, plasmonic, and excitonic features shape the excitation spectra of ZnO obtained by EELS. In particular, excitonic contributions~\cite{Mang1995SSC,Teke2004PRB,Wu2012Plasmonics,Gori2010PRB} have been found to be responsible for distinct low-energy features in the response properties of ZnO, including the very sharp onset of the EELS spectra~\cite{Huang2011PRB,Wu2012Plasmonics,Granerod2018PRB,Granerod2018JAP,Yeh2021PRApp}, and the significant temperature dependence of the optical gaps~\cite{Liang1968PRL,Mang1995SSC,Adachi2004handbook,Teke2004PRB,Tsoi2006PRB,Granerod2018JAP}. 
Other features at larger energies have been associated with single-particle and plasmonic excitations~\cite{Dorn1977PRB,Wang2005APL,Zhang2006Micron,Wang2007Micron,Yeh2021PRApp}, described with optical calculations in both, the independent particle (IPA) and the random phase (RPA) approximations~\cite{Zhang2006Micron,Laskowski2006PRB,Gori2010PRB,Riefer2017JPCM,Yeh2021PRApp}. The important role of electron-hole interactions in the absorption spectra of ZnO has been studied with the
Bethe-Salpeter equation (BSE) method~\cite{Laskowski2006PRB,Schleife2009PRB,Gori2010PRB,Dvorak2013PRL,Riefer2017JPCM,Zhang2018PRB,Shen2022APL}, showing results in excellent agreement with e.g. ellipsometry measurements~\cite{Gori2010PRB}.

Another distinct feature of the dielectric response of ZnO is its strong anisotropy along the direction perpendicular to the hexagonal plane of its wurzite crystalline structure. In particular, the static dielectric constant of ZnO is about 10\% smaller in this direction~\cite{Adachi2013optical}. 
The high-symmetry lines in the Brillouin zone are therefore classified as in-plane and out-of-plane, according to its component along the perpendicular direction. 
The adsorption spectrum in the optical limit, $\mathbf{q}\to 0$, along in-plane and out-of-plane directions also differs at finite energies, as found both experimentally and theoretically by changing the polarization of the external electric field~\cite{RakelIBOOK2008,Gori2010PRB,Schleife2009PRB,Yeh2021PRApp}. 
In addition, momentum-resolved EELS experiments have been carried out~\cite{Wang2007Micron} to describe how the loss function disperses along the $[120]$ in-plane and $[001]$ out-of-plane crystalline directions. However, the effects of the anisotropy at finite values of $\mathbf{q}$ have not yet been addressed in the literature.

In this work, we conduct {\it ab-initio} electronic structure and response function calculations combined with momentum-resolved EELS experiments, to examine the role of different physical effects on the momentum and energy-dependent response functions of ZnO, including a theoretical description of the anisotropy at finite $\mathbf{q}$. Several theoretical approaches are used, allowing us to systematically unravel the key many-body features in the spectra. We perform DFT calculations with standard and hybrid functionals, including DFT$+U$. We then benchmark optical calculations with the different DFT inputs at the IPA and RPA levels. The DFT$+U$ scheme, which results in a better agreement with the EELS spectra, is then chosen to study the excitonic contributions at the BSE level. 

The paper is organized as follows:
Sec.~\ref{sec:theory} introduces the theory and main concepts behind EELS experiments and first-principles calculations at the different levels of approximation. 
Sec.~\ref{sec:methods} provides experimental and computational details. In Sec.~\ref{sec:results} we describe and analyze the main results, and finally, Sec.~\ref{sec:conclusions} holds the conclusions. 

\section{Theory}
\label{sec:theory}
In EELS experiments, the double-differential
scattering cross-section, $\sigma$, per unit of volume resolved in energy $\omega$ and momentum $\mathbf{Q}$, proportional to the solid angle $\varphi$, is given by (in atomic units)~\cite{Egerton2011book,Alkauskas2013PRB}:
\begin{equation}
\frac{1}{V} \frac{d^2\sigma}{d\varphi d\omega} = \frac{\gamma^2}{2 \pi^2} \frac{k_f}{k_i} v(\mathbf{Q}) L(\mathbf{Q},\omega),
\label{eq:cross_sec}
\end{equation}
where $\gamma=1/\sqrt{1-v_e^2/c^2}$ is the relativistic factor stemming from the velocity $v_e$ of the incident electrons; 
$\mathbf{k_i}$ and $\mathbf{k_f}$ are the initial and the final momenta of the electrons with $k_i \equiv |\mathbf{k_i}|$ and $k_f \equiv |\mathbf{k_f}|$, so that the transferred momentum is given by $\mathbf{Q}=\mathbf{k_f}-\mathbf{k_i}$, with $Q \equiv |\mathbf{Q}|$; while $v(\mathbf{Q})=4\pi/Q^2$ is the Coulomb potential in reciprocal space. 
The loss function is defined by the imaginary part of the macroscopic inverse dielectric function, $L(\mathbf{Q},\omega)\equiv \ip[-\varepsilon^{-1}_M(\mathbf{Q},\omega)]$.
The macroscopic inverse dielectric function, $\epsilon^{-1}_M$, is related to the macroscopic polarizability, $\chi_M$,  by:
\begin{equation}
\varepsilon^{-1}_M (\mathbf{Q},\omega) = 1 + v(\mathbf{Q}) \chi_M (\mathbf{Q},\omega).
\end{equation}

Note that the macroscopic momentum transfer, $\mathbf{Q}$, can exceed the first Brillouin zone. The macroscopic and microscopic density-density response functions are related by
\begin{equation}
    \chi_M (\mathbf{Q},\omega) = \chi_{\mathbf{G G}} (\mathbf{q},\omega),
\end{equation}
where $\mathbf{Q}=\mathbf{q}+\mathbf{G}$, $\mathbf{q}$ is confined to the 
first Brillouin zone, and $\mathbf{G}$ is a reciprocal lattice vector. $\chi_{\mathbf{G G'}} (\mathbf{q},\omega)$ is the Fourier transform in space and time of the non-local dynamic polarizability function $\chi(\mathbf{r},\mathbf{r'},t-t')$. 

In the many-body theory, the microscopic polarizability can be evaluated perturbatively from Kohn-Sham DFT by solving the Dyson equation for this operator:
\begin{equation}
    \chi = \chi_0 + \chi K \chi_0, 
    \label{eq:dyson}
\end{equation}
where for simplicity the dependencies, $\mathbf{G G'}$ and $(\mathbf{q},\omega)$, have been omitted, $\chi_0$ is the independent-particle polarizability, and the kernel $K$ defines the given level of theory. At the IPA level, $K_{\text{IPA}}=0$, only single-particle transitions are accounted for, while local field effects are included in RPA, $K_{\text{RPA}} = v(\mathbf{q + G})$. 
Electron-hole interactions can also be included in Eq.~\eqref{eq:dyson} by using four-particle operators, i.e. replacing the polarizabilities (two-particle operators) by four-particle response functions and using the Bethe-Salpeter kernel, $K_{\text{BSE}} = v(\mathbf{q + G}) - W_{\mathbf{G G'}} (\mathbf{q}, \omega)$, where $W$ is the dynamically screened Coulomb potential. In practice, the dynamical effects are neglected by considering only the $W_{\mathbf{G G'}} (\mathbf{q}, \omega=0)$ term at the RPA level ($W = v + v \chi v$), leading to a simplified Bethe-Salpeter equation with a static kernel~\cite{martin2016book}.

In this work, we compute the IPA polarizability $\chi_0$ from Kohn-Sham (KS) eigenvalues and eigenvectors as
\begin{multline}
    \chi_{0\mathbf{G G'}}(\mathbf{q},\omega) = 2 \sum_{v,c} \int_{BZ} \frac{d \mathbf{k}}{(2\pi)^3} \rho^*_{v c \mathbf{k}}(\mathbf{q},\mathbf{G}) \rho_{v c \mathbf{k}}(\mathbf{q},\mathbf{G'}) \times \\
     f_{v \mathbf{k-q}}(1-f_{c \mathbf{k}})
    \left [ \frac{2 \Omega^{\text{KS}}_{v c \mathbf{k} \mathbf{q}}}{\omega^2- (\Omega^{\text{KS}}_{v c \mathbf{k} \mathbf{q}})^2} \right ],
    \label{eq:Xipa}
\end{multline}
where the factor $2$ accounts for the spin degeneracy, the $v$ and $c$ indices run respectively over valence and conduction bands, $\rho_{v c \mathbf{k}}(\mathbf{q},\mathbf{G}) \equiv \langle v \mathbf{k} | e^{i (\mathbf{q} + \mathbf{G}) \cdot \mathbf{r}} | c \mathbf{k-q} \rangle$ are transition matrix elements, the $f$ factors are the occupations of the KS states, $\Omega^{\text{KS}}_{v c \mathbf{k} \mathbf{q}} = (\epsilon_{c \mathbf{k}}-\epsilon_{v \mathbf{k-q}}) -i\delta$ are KS single-particle transitions, and the limit $\delta \to 0^+$ is implicit and ensures the correct time ordering~\cite{martin2016book}.

\section{Methods}
\label{sec:methods}
To make a reliable comparison between EELS measurements and the theoretical loss function, the QP band structure used as input for the computation of the spectra needs to be in good agreement with the experimental band structure and the corresponding density of states, such as obtained from angle-resolved photo-emission spectroscopy (ARPES)~\cite{Lim2012PRB} and X-ray photo-electron spectroscopy (XPS)~\cite{Preston2008PRB,King2009PRB}.
However, as discussed in the introduction, computationally obtaining a QP band structure of ZnO in good agreement with the experiment is in itself challenging. 
We therefore compute the DFT band structure and density of states at different levels of theory, as will be discussed in Sec.~\ref{sec:com_dets}. These results are compared with the experimental data of Refs.~\cite{Lim2012PRB,King2009PRB}, focusing on the role of orbital hybridization in the position of the Zn-$d$ states. Such benchmarkings proved to be important for determining a suitable starting point to compute the response functions of ZnO at the different IPA, RPA and BSE levels of theory.

A precise theory-experiment comparison of the loss function also requires careful processing of the EELS measurements. 
The recorded intensity is proportional to the differential cross-section defined in Eq.~\eqref{eq:cross_sec}; however, because of the finite momentum and energy resolutions the resulting spectra are convoluted in these two variables, which hinders the factorization of the loss function, $L(\mathbf{q}, \omega)$. 
Furthermore, the measured intensity is sensitive to several parameters related to the specific experiment, such as instrumental broadening, exposure time, and thickness variations of the sample. 
It is therefore common to model the absolute thickness and perform an approximate normalization using e.g., the method known as Kramers-Kronig sum rule~\cite{Egerton2011book}. 
While such procedures can be sufficient for analyzing trends in peak positions, they neglect broadening, surface-mode scattering, and retardation effects~\cite{Egerton2011book}, and thus may be insufficient to analyze the dispersion of the intensity with $\mathbf{q}$. 
Hence, we use a simple method based on the physical constraint of the $f$-sum rule~\cite{Egerton2011book,martin2016book} to extract the experimental loss function, as detailed in App.~\ref{sec:loss_exp}. We also perform Kramers-Kronig analysis with the procedure described in App.~\ref{sec:KK_exp} to obtain the momentum-dependent absorption spectra from the experimental loss function. Similar methods have also been used for inelastic X-ray scattering spectroscopy (IXSS) measurements~\cite{Schulke1988PRB,Schulke1989PRB,Schulke1995PRB,Watanabe2006BCSJ,Weissker2010PRB}.

In this work, we use transferred momenta constrained to the first Brillouin zone ($\mathbf{Q}=\mathbf{q}$), for both experiments and {\it ab-initio} calculations. Both theoretical and experimental data are given in coarse grids along different directions in $\mathbf{q}$-space, while the frequency grid is denser with respect to the energy range we consider. Thus, first-order spline interpolation is used to obtain the smooth color-maps in Secs.~\ref{sec:response} and \ref{sec:excitons}. 

\subsection{Experimental details}
\label{sec:exp_details}
Single crystal ZnO with at least $99.99\%$ purity was purchased from the MTI  Corporation. Transmission {e}lectron {m}icroscopy (TEM) specimens were prepared by mechanical grinding using an Allied Multiprep system and polished with a Gatan Precision Ion Polishing System (PIPS) II. 
Two different samples were prepared to access both the in-plane and out-of-plane directions. 

Momentum-resolved EELS measurements were obtained in TEM mode, using a Gatan GIF Quantum 965 spectrometer attached to a monochromated FEI Titan G2 60-300, operated at $60$~keV. We used the $\omega$-$\mathbf q$ mapping technique, where an angle-selection slit was placed along the $\Gamma M$, $\Gamma K$, $MK$, and $\Gamma A$ high-symmetry lines in the diffraction plane. The spectrometer disperses the electrons according to their velocity perpendicular to the direction of the slit, and the resulting $\omega$-$\mathbf q$ map is a 2D intensity distribution as a function of energy loss (eV) and momentum $\mathbf{q}$ (\AA$^{-1}$).

The momentum resolution is estimated by the effective width of the angle-selecting slit to be  $0.6$~\AA$^{-1}$ in the in-plane directions and $0.45$~\AA$^{-1}$ in the out-of-plane one, as given by the angular range used in each case. The energy resolution is estimated as $\lesssim 0.30$~eV, quantified by the full-width at half-maximum of the zero-loss peak (ZLP), consisting of electrons elastically scattered or transmitted with energy losses below the resolution of the instrumentation. The energy dispersion of the spectrometer was set to $0.025$~eV/channel, as a compromise between energy resolution and signal intensity.

At the $\Gamma$ points ($\mathbf{q} \to 0$), the ZLP is dominated by quasi-elastic scattering, while the relative contribution of inelastic scattering increases at finite $\mathbf{q}$. The finite size of Bragg spots around $\Gamma$ constrains the accessibility to the $\mathbf{q} \to 0$ limit. Moreover, the dynamic range of the detector does not provide a good signal-to-noise ratio for the full $\mathbf{q}$-range of interest. Therefore, the data acquisition was split in two $\mathbf{q}$-ranges to optimize the quality of the signal: $q_{\text{in}} \leq 0.15$~\AA$^{-1}$ and $q_{\text{in}} \geq 0.20$~\AA$^{-1}$ in the in-plane directions, and $q_{\text{out}} \leq 0.30$~\AA$^{-1}$ and $q_{\text{out}} \geq 0.35$~\AA$^{-1}$ in the out-of-plane one.

The individual EELS spectra were extracted from the full data sets
by summing few adjacent spectra in a maximum range of $\sim0.1$~\AA$^{-1}$, for statistical averaging. The resulting spectra were then extracted in intervals of $0.05$~\AA$^{-1}$. 
The background from the ZLP was subtracted using the standard power law model described in Ref.~\cite{Stoger2008Micron}. We thereafter have applied a Savitzky-Golay filter to remove remaining fluctuations in the spectra (see details in Sec.~IV of the \suppinfo).

\subsection{Computational details}
\label{sec:com_dets}
DFT calculations were performed using the plane wave implementation of the {\qe} package~\cite{QE1,QE2} with the Perdew-Burke-Ernzerhof (PBE) variant of the GGA  functional~\cite{Perdew1996PRL}, its hybrid (tuned) PBE0 version~\cite{Perdew1996JCP}, and PBE in a DFT$+U$ scheme with tunable $U$ parameters~\cite{Cococcioni2005PRB,Campo2010JPCM}. 
We adopted the norm-conserving optimized Vanderbilt pseudopotentials of Ref.~\cite{Hamann2013PRB}, with a kinetic energy cutoff of $70$~Ry for the wave-functions.
ZnO is modeled in its hexagonal wurtzite crystal structure with lattice constants set to the experimental values, $a=3.25$~\AA{} and $c=5.21$~\AA~\cite{Adachi2004handbook,klingshirn2010zinc}.
In all the DFT calculations, the Brillouin zone was sampled with a $36\times36\times24$ Monkhorst-Pack grid. 
Notice that even if the DFT calculations are over-converged with respect to this grid, such dense grids are necessary to obtain a smooth representation of the response functions.

For hybrid PBE0-type calculations, the exact exchange fraction is set to 0.22, slightly lower than the default value of 0.25, selected to reproduce the experimental bandgap at room temperature. 
Similarly, for PBE$+U$ calculations, $U$ values for O and Zn atoms were set to $U_{\text{O}}=7.5$~eV and $U_{\text{Zn}}=9.5$~eV. This combination reproduces both the experimental bandgap and the position of the $d$-states obtained from ARPES and XPS measurements in Refs.~\cite{Lim2012PRB,King2009PRB}.  

One-shot $G_0W_0$ and QP self-consistent $GW_0$ and $GW$ calculations starting from PBE results were performed with the {\yambo} code~\cite{Marini2009CPC,Sangalli2019JPCM}. We concluded that the QP band structure within these approaches can be approximately obtained from the PBE results by applying scissors and stretching operations~\cite{martin2016book}, and thus be used to compute the direct and inverse dielectric functions. 
The details of such $GW$ calculations are provided in Sec.~I of the \suppinfo.

The calculations of the optical spectra within IPA, RPA, and BSE were also performed with {\yambo}, including the use of scissors operators~\cite{delSole1993PRB}. The Brillouin zone for both $\mathbf{k}$ and $\mathbf{q}$ spaces was sampled with the same grid as DFT, except for the BSE calculations, where a smaller grid of $18\times18\times12$ $\mathbf{k}$-points was adopted due to their high computational cost. The IPA and RPA polarizability matrices were computed using a total number of occupied plus unoccupied states of 100 and an energy cutoff of $5$~Ry for the reciprocal lattice vectors. For the BSE calculations, we use a static $W$ matrix with 7 occupied and 14 unoccupied states respectively below and above the Fermi level, and the same cutoff of $5$~Ry. The BSE kernel is constructed directly from the PBE0 and PBE$+U$ eigenvalues and eigenstates rather than the more common choice of using $GW$, since the bandgap is already accurate within those approaches. All the spectra are computed with a Lorentzian broadening of $0.2$~eV.

\section{Results}
\label{sec:results}
The results of our work are presented in four subsections. Sec.~\ref{sec:start} describes DFT results obtained at the PBE, PBE0, and PBE$+U$ levels of theory, and the role of orbital hybridization in the band structure of ZnO. Its importance for interpreting the main features in the loss function is highlighted in Sec.~\ref{sec:response}, by comparing our theoretical and experimental spectra with previous studies in the optical limit. 
In Sec.~\ref{sec:q_response} we provide a more detailed description of the direct and inverse dielectric functions, as obtained at finite momentum along different directions in the Brillouin zone. The in-plane $\Gamma M$ line, which has the most accurate experimental data, is used to benchmark the different IPA and RPA approaches, and select the best input for BSE calculations. We discuss the dispersion of the main features defined in the optical limit, including single-particle and plasmonic excitations and the strong anisotropy along the out-of-plane direction. Finally, Sec.~\ref{sec:excitons} is dedicated to excitonic effects.

To facilitate the following descriptions, here we introduce some useful concepts: the peaks in $\ip[-\varepsilon^{-1}]$ and $\ip[\varepsilon]$ will be called two-body or many-body states, or simply excitation states. 
For instance, such peaks could be originated from single-particle transitions from valence to conduction states, or have plasmonic or excitonic character. The dispersion of such excitation states with transferred momentum, $\mathbf{q}$, defines a band structure of excitations in $\mathbf{q}$-space, for both $\ip[-\varepsilon^{-1}]$ and $\ip[\varepsilon]$, analogous to the standard band structure of valence and conduction (one-body) states in $\mathbf{k}$-space, 
like the plot in panel a) of Fig.~\ref{fig:ZnO_bs_dos}.
While one-body-states from valence and conduction  respectively have negative and positive energy with respect to the Fermi level, the poles of the response functions corresponding to the (many-body) excitation states are positively defined (see e.g. Eq.~\eqref{eq:Xipa}).  
Another property of the one-body-state band structure is the possibility to have bands crossing at some point in the Brillouin zone, changing the order of the bands before and after a crossing. Due to the band reordering, similar crossings can happen for the excitation states, not necessarily at the same points.

\subsection{Band structure and electronic density}
\label{sec:start}
\begin{figure*}[t!]
  \includegraphics[width=1.00\textwidth]{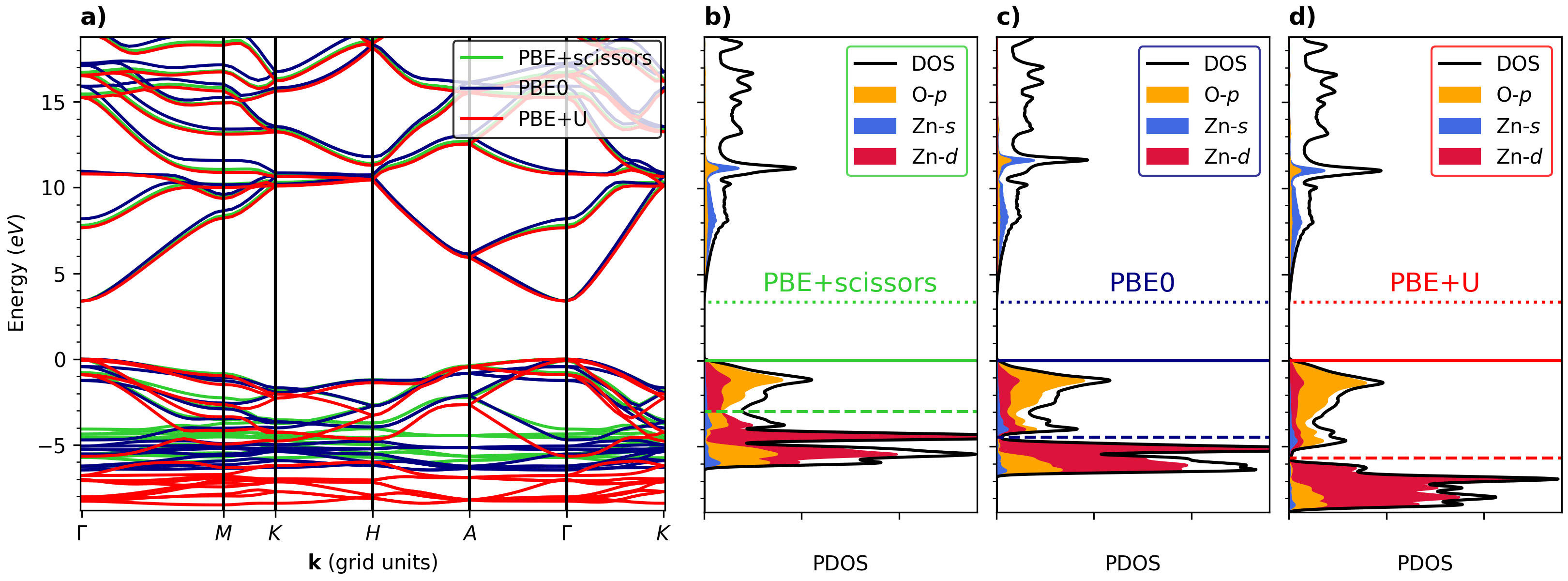}
    \caption{Band structure (panel a) and projected density of state (PDOS) (panels b-d) of ZnO obtained with PBE, PBE0 and PBE$+U$. Scissors corrections of $2.51$~eV and $0.02$~eV  have been applied respectively to PBE and PBE0 results in order to match the PBE$+U$ gap of $3.38$~eV.
    The momentum $\mathbf k$ is given in grid units, i.e. the length of the $\Gamma M$, $M K$, $K H$, $H A$, $A \Gamma$ and $\Gamma K$ high-symmetry lines in the band structure is proportional to the number of sampling points along them, which is 19 for $\Gamma M$, 7 for $M K$, and 13 for the rest; with the $36\times36\times24$ $\mathbf k$-grid.
    Three horizontal lines are drawn in each of the PDOS plots, the bottom of the conduction band determining the gap (dotted), the Fermi level at null energy (solid), and the energy from which the DOS become predominantly of $d$-character (dashed); $-2.96$~eV for PBE, $-4.45$~eV for PBE0 and $-5.66$~eV for PBE$+U$.  
    }
    \label{fig:ZnO_bs_dos}
\end{figure*}
Fig.~\ref{fig:ZnO_bs_dos} compares the KS band structure (panel a) and the projected density of states (PDOS) (panels b-d) of ZnO computed with PBE, PBE0, and PBE$+U$. We obtain a fundamental gap of $0.87$~eV for PBE, $3.36$~eV for PBE0 and $3.38$~eV for PBE$+U$ with the chosen parameters.  
Here, scissors corrections of $2.51$~eV and $0.02$~eV respectively, align the PBE and PBE0 band structures to the PBE$+U$ gap.
The comparison shows very similar conduction bands in the three approaches, except for a slight stretching of the PBE0 bands compared to those of PBE and PBE$+U$.
However, the valence bands differ considerably, in particular, the positions of the $d$-states (located at the bottom part of the figure) of PBE and PBE0 bands are too high in energy compared to those of PBE$+U$ and the ARPES experiments of Ref.~\cite{Lim2012PRB}. 
In the PDOS plots (panels b-c), the horizontal dashed lines indicate a change from O-$p$ to Zn-$d$ orbital character that predominates in the total density of states (DOS). 
The change occurs at $-2.96$~eV, $-4.45$~eV, and $-5.66$~eV, respectively for PBE, PBE0, and PBE$+U$. 
Among these, the PBE$+U$ DOS is the most consistent one with the ARPES and XPS measurements of Refs.~\cite{Lim2012PRB,King2009PRB}, as by construction its maximum weight is located around the experimental value of $-7.5$~eV.

Perturbative approaches, whether simple schemes such as correcting KS energies with scissors and stretching operations, or advanced methods such as one-shot or QP self-consistent $GW$, rest on the assumption that DFT with standard functionals provides quantitatively accurate electronic orbitals, and associated electronic density, but this is not the case for ZnO~\cite{Usuda2002PRB,Lim2012PRB}.
Alternatively, fixing the bandgap in a (fully) self-consistent DFT approach can be done with a hybrid functional like PBE0, by tuning the fraction of exact exchange, but it does not provide much improvement for the position of the $d$-states, even combined with QP $GW$ methods, as also found e.g. in Ref.~\cite{Lim2012PRB}.

Our PBE$+U$ PDOS results confirm the importance of having a correct orbital hybridization. They also show that by tuning the $U$ parameters it is possible to improve the description of both features, the gap and the $d$-states, as compared to ARPES~\cite{Lim2012PRB} and XPS experiments~\cite{Preston2008PRB,King2009PRB}. There is a plethora of possible values of $U$ used for ZnO in the literature~\cite{Harun2020ResInPhys}, even formulated as pseudo-hybrid functionals~\cite{Agapito2015PRX,Gopal2015PRB}. We chose the combination, as mentioned in Sec.~\ref{sec:com_dets}, that best agrees with the experimental band structure and DOS of Refs.~\cite{Lim2012PRB,King2009PRB}, and the gap of $3.37$~eV at room temperature~\cite{Mang1995SSC,Harun2020ResInPhys}. This choice reduces the variety of possible response functions that can be found with different values of $U$ (see for example Fig. 7 of Ref.~\cite{Harun2020ResInPhys}).

 \begin{figure}   
  \includegraphics[width=0.49\textwidth]{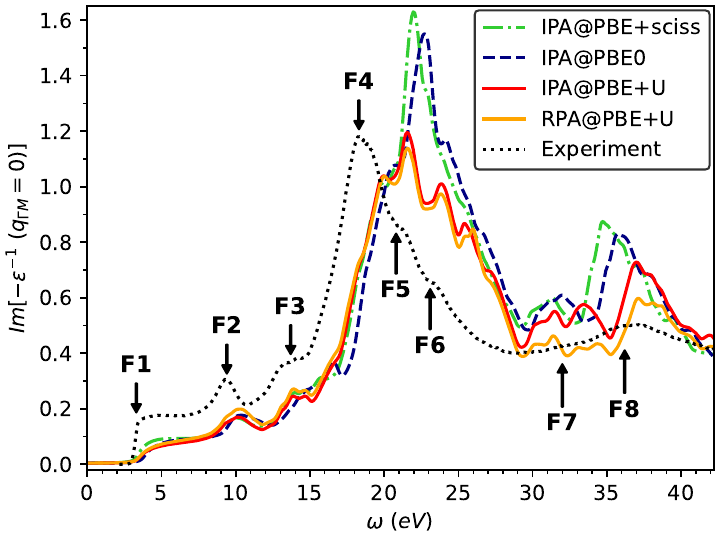}
    \caption{Loss function of ZnO in the $\mathbf{q}\to 0$ limit along the $\Gamma M$ line computed at the IPA level with PBE, PBE0 and PBE$+U$ and the RPA level on top of PBE$+U$. The normalized experimental measurements are also plotted. The same scissors corrections from Fig.~\ref{fig:ZnO_bs_dos} are used. Eight features of the loss function are indicated in the plot.} 
    \label{fig:ZnO_loss}
\end{figure}

\subsection{Response properties in the optical limit}
\label{sec:response}
Fig.~\ref{fig:ZnO_loss} compares the computed loss function, $L(\mathbf{q}, \omega)$, of ZnO in the $\mathbf{q}\to 0$ limit along the in-plane $\Gamma M$ line; obtained at the IPA level on top of scissor-corrected PBE, PBE0, and PBE$+U$; with the corresponding EELS measurements. For PBE$+U$, the spectrum at the RPA level is also shown. In the figure, we indicate eight main features of $L(q_{\Gamma M}\to 0, \omega)$, labeled as {\bf F1}, {\bf F2}, ..., {\bf F8}, following a similar notation to Refs.~\cite{Zhang2006Micron,Wu2012Plasmonics}. 
A detailed comparison of the experimental and theoretical features allows us to asses the accuracy of the different levels of theory, and facilitates the interpretation of the features with respect to previous studies.

The experimental loss function exhibits a sharp onset at {\bf F1}, with an optical gap of $3.2$~eV, estimated from standard parabolic fitting used in EELS~\cite{Zhan2018SciRep}. 
This value is consistent with the range of values reported in other experiments at room temperature~\cite{Mang1995SSC,Jellison1998PRB,Srikant1998JAP,Zhan2018SciRep,Granerod2018JAP}. However, the fitted bandgap is around $0.1$~eV smaller than the reference value of $3.37$~eV from optical measurements~\cite{Thomas1960JPCS,Mang1995SSC,Jellison1998PRB,Teke2004PRB}, taking an excitonic binding energy of $60$~meV into account~\cite{Thomas1960JPCS,Jellison1998PRB,Muth1999JAP,Teke2004PRB}. This underestimation is also common in other spectroscopies susceptible to valence band-donor transitions from the bulk of the system, when comparing to experimental techniques probing the surface, such as ellipsometry and luminescence~\cite{Srikant1998JAP,Teke2004PRB}.

After the onset, there is a prominent peak at $9.4$~eV~({\bf F2}) and a shoulder structure between $12$ and $15$~eV~({\bf F3}). At larger energies we find the bulk plasmon, with a double peak located at $18.4$~eV and $18.9$~eV ({\bf F4}), followed by two shoulders around $20.8$~eV ({\bf F5}) and $23.1$~eV ({\bf F6}). Finally, there are two shallow hills, the first one around $32$~eV ({\bf F7}), which vanishes in the  $\mathbf{q} \to 0$ limit, and a more prominent one around $36$~eV ({\bf F8}) showing only small variations with $\mathbf{q}$.
The measured energy-loss spectrum is in very good agreement with previous EELS experiments~\cite{Hengehold1970PRB,Wang2005APL,Zhang2006Micron,Wang2007Micron,Huang2011PRB,Wu2012Plasmonics,Yeh2021PRApp}, according to their energy scale and resolution. While many of these measurements were performed for different ZnO nanostructures, where surface plasmons may be present, the rest of the features correspond to those from bulk ZnO~\cite{Wang2005APL,Zhang2006Micron,Wang2007Micron,Wu2012Plasmonics,Yeh2021PRApp}.

On the theoretical side, the onset {\bf F1} is smoother and lower in intensity, mainly due to the absence of electron-hole interactions in the IPA and RPA approximations. The position of the {\bf F2} peak is overestimated by $0.8$~eV, $0.5$~eV, and $0.7$~eV, for PBE, PBE0, and PBE$+U$, respectively. The agreement in the PBE0 case is slightly better due to the asymmetry caused by the superposition of two peaks, absent in the other theories and the experiment. 
For all the four types of calculations, 
the shoulder {\bf F3} arises from a set of peaks. In the case of PBE and PBE0, the width of the {\bf F3} region is larger, while PBE$+U$ agrees better with the experiment and shows better-defined peaks. 
The intensity of {\bf F2} and {\bf F3} is also lower in the theory, which is linked to the overestimation of the position of the plasmon and its tail. 
The theoretical spectra also exhibit an extra subtle shoulder around $18.3$~eV for PBE and PBE$+U$, and $18.8$~eV for PBE0, which is not present in the experiment.

As in the experiment, the plasmon presents a double peak shape, with maximums located at $19.8$~eV and $22.0$~eV for PBE, $20.6$~eV and $22.7$~eV for PBE0, and $19.9$~eV and $21.6$~eV for PBE$+U$, overestimating the experimental value on $3.1$~eV, $3.8$~eV, and $2.7$~eV, respectively. 
In all the theoretical approaches, the intensity is larger for the second peak, pronouncedly so for PBE and PBE0. In general, the relative intensity among the different features in PBE$+U$ is in much better agreement with the experiment, due to a better description of {\bf F4}. 
Moreover, whereas {\bf F5} and {\bf F6} are subtle shoulders in the PBE and PBE0 spectra, in PBE$+U$ they are prominent peaks at $23.9$~eV and $25.6$~eV respectively. 

Last, the agreement between experiment and theory is very good for the {\bf F7} and {\bf F8} hills. The position of {\bf F7} is around $30.9$~eV for PBE, $31.8$~eV for PBE0, and $32.3$~eV for PBE$+U$, the latter being more structured. {\bf F8} is located around $34.7$~eV, $35.8$~eV and $37.0$~eV, for PBE, PBE0, and PBE$+U$, respectively. The intensities of these peaks are also affected by the position and tail of the plasmon, which are higher for PBE and PBE0 than for PBE$+U$. In particular, the RPA@PBE$+U$ intensity of {\bf F7} tends to vanish in the $\mathbf{q} \to 0$ like in the experiment, as described in the next section.

While the theory overestimates the position of the plasmon and there are several other smaller shifts for the rest of the features, the overall agreement between theory and experiment is quite good. Except for some contraction of the spectrum beyond $9~$eV, the overall shape of scissor-corrected PBE is very similar to that of PBE0,  which can be attributed to a similar orbital hybridization with some stretching of the PBE0 bands compared to PBE, as discussed in Sec.~\ref{sec:start}. In contrast, PBE$+U$, with its improved description of the $d$-states gives a spectrum much closer to the experiment, even if a significant shift toward larger energies remains. 

Comparing the IPA and RPA results in Fig.~\ref{fig:ZnO_loss} allows us to assess the role of local-field effects, as obtained by solving the Dyson equation in Eq.~\eqref{eq:dyson} at this level of theory.
These effects broaden the tail of the plasmon, due to long-range Coulomb interactions, renormalizing the intensity of the peaks in the spectra, even at energies below the plasmon such as the {\bf F2} feature. As shown in Fig.~\ref{fig:ZnO_loss}, the renormalization at high energies is quite large, even for the macroscopic inverse dielectric function. This is the main reason for the cumbersome convergence of $GW$ calculations, requiring the evaluation of the microscopic polarizability with a high energy cutoff for the lattice vectors and the number of empty states~\cite{Shih2010PRL,Friedrich2011PRB,Rangel2020CPC}.

A better understanding of the main orbital character of the features in the loss function of ZnO, can be obtained by analyzing the dipole matrix elements and identifying the bands responsible for the main electronic transitions that form the given features. Such an analysis is provided in Sec.~II of the \suppinfo, at the IPA@PBE$+U$ level of theory. Overall, we find that the compositions of all these features is consistent with previous works~\cite{Hengehold1970PRB,Zhang2006Micron,Wang2007Micron,Yeh2021PRApp}, although we address the main disagreements for {\bf F2} and {\bf F3}. 

According to our results, {\bf F1} and {\bf F2} are dominated by excitations from O-$p$ states, while previous analyses based on DFT calculations with a GGA-based functional~\cite{Yeh2021PRApp} assign a Zn-$d$ character to {\bf F2}, which could be due to an incorrect orbital hybridization similar to our PBE results. 
Instead, an O-$p$ character is assigned in Ref.~\cite{Zhang2006Micron}. Moreover, instead of a structured shoulder shape at {\bf F3}, some of the earlier experiments have reported only 1-2 peaks in this region~\cite{Hengehold1970PRB,Zhang2006Micron,Wu2012Plasmonics}, which we attribute to limitations in the resolution of their measurements. 
Consequently, different characters have been assigned to {\bf F3}. 
The literature~\cite{Zhang2006Micron,Wu2012Plasmonics,Yeh2021PRApp} agrees on one of the peaks around $13.5~$eV corresponding to transitions from Zn-$d$ states. A surface plasmon is reported at $15.8~$eV only in Ref.~\cite{Wu2012Plasmonics}. Three peaks are identified in Ref.~\cite{Yeh2021PRApp} at $13.0~$eV, $13.5~$eV and $14.8~$eV, respectively with Zn-$d$, O-$p$ and mixed character, this is more consistent with the mixed character of {\bf F3} in our results. 
An orientational dependent part of {\bf F3}~\cite{Wang2007Micron,Yeh2021PRApp} and a momentum-dependent composition at finite $\mathbf{q}$ has also been reported~\cite{Wang2007Micron}. 

A double-peaked plasmon has not been previously reported in Refs.~\cite{Zhang2006Micron,Wang2007Micron,Wu2012Plasmonics,Yeh2021PRApp}, however, the asymmetric shape of the plasmon peak measured in Ref.~\cite{Yeh2021PRApp} is consistent with our findings. 
The plasmon and the rest of the {\bf F5-8} features are also expected to have mixed characters, although they have been less studied in the literature. Furthermore, as detailed in Sec.~II of the \suppinfo, we find that the weight of the different orbital contributions is not the same for all the features, for example, {\bf F5} and {\bf F6} have a dominant Zn-$d$ component. 
The comparison with the experiment in Fig.~\ref{fig:ZnO_loss} shows the limitations of the theory for features at such large energies. Therefore, a better agreement is expected with an improved description of the $d$-states beyond DFT$+U$.

\subsection{Inverse and direct dielectric functions in $\mathbf{q}$-space}
\label{sec:q_response}
\begin{figure*}
    \centering    \includegraphics[width=0.98\textwidth]{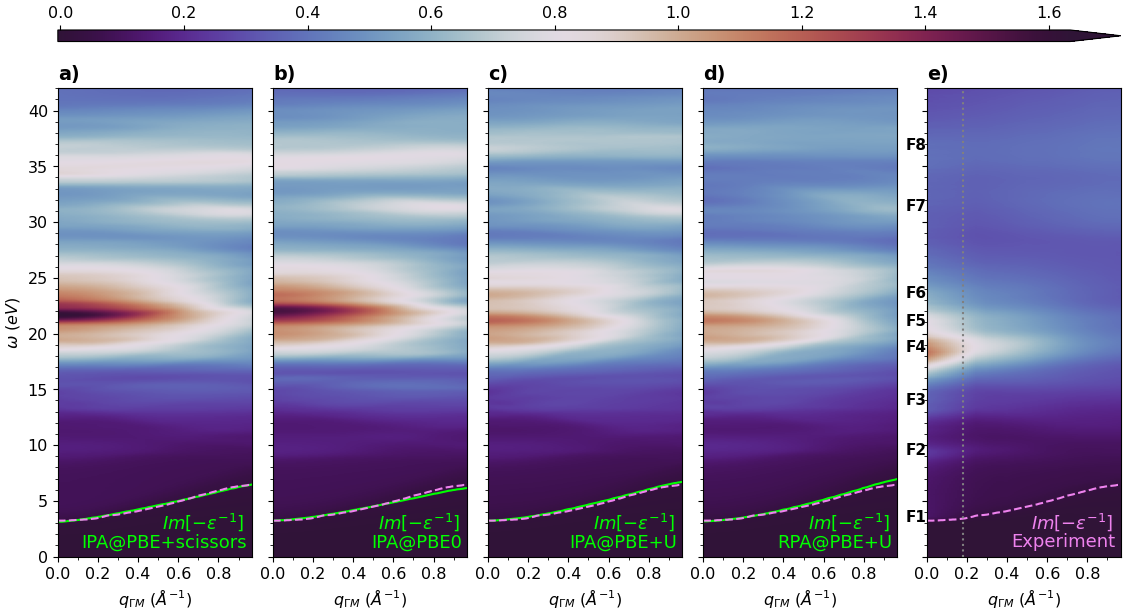}
    \caption{Color-map representation of the energy and momentum-dependent inverse dielectric function, $\ip[-\varepsilon^{-1}(\mathbf{q},\omega)]$, along the $\Gamma M$ line. Comparison between several flavors of theory: IPA@PBE (a), IPA@PBE0 (b), IPA@PBE$+U$ (c) and RPA@PBE$+U$ (d), and EELS experiments (e). The dispersion of the features {\bf F1-4} defined in Fig.~\ref{fig:ZnO_loss} is appreciable, as indicated with labels in the experimental panel (e). The edge of the onsets is drawn with approximate lines to guide the eye, while a fit is provided in Table~\ref{tab:gap_mass}. The dotted gray line at $q_{\Gamma M}=0.18$~\AA$^{-1}$ in panel e) indicates the change of experimental data set (see description in Sec.~\ref{sec:exp_details}).}
    \label{fig:cmap_GM_eel}
\end{figure*}

In this section, we address several properties of the momentum dependence of the direct and inverse response functions of ZnO. Even if PBE$+U$ results are closer to the experiment in the $\mathbf{q}\to 0$ limit, it is not obvious whether this simple approach provides accurate results also at finite momentum. Therefore, in Fig.~\ref{fig:cmap_GM_eel} we benchmark the dispersion of the features obtained with the same four theoretical approaches and the experiment of Fig.~\ref{fig:ZnO_loss}, along the $\Gamma M$ line. 
The experimental momentum-dependent loss function was obtained with the method described in App.~\ref{sec:loss_exp}. The comparison between theory and experiment shows the same level of agreement for the whole $\Gamma M$ line.

\begin{table}
\centering
\begin{ruledtabular}
\begin{tabular}{lccc}
\\[-3pt]
           &  $c_1 ($\AA$)$             & $c_2 ($\AA$^2)$             & $c_3  ($\AA$^3)$             \\[5pt]
\hline\\[-3pt]
IPA@PBE    & $0.61 \pm 0.01$   & $1.95 \pm 0.06$   & $-2.62 \pm 0.13$  \\
IPA@PBE0   & $0.33 \pm 0.01$   & $2.58 \pm 0.06$   & $-4.00 \pm 0.12$  \\
IPA@PBE$+U$  & $0.22 \pm 0.01$   & $3.60 \pm 0.06$   & $-5.29 \pm 0.12$  \\
RPA@PBE$+U$  & $0.22 \pm 0.01$   & $3.64 \pm 0.06$   & $-4.65 \pm 0.13$  \\
BSE@PBE$+U$  & $0.004 \pm 0.11$   & $3.67 \pm 0.22$   & $-4.16 \pm 0.47$  \\
EELS       & $0.01 \pm 0.05$   & $4.13 \pm 0.20$   & $-6.38 \pm 0.41$  \\
\end{tabular}
\end{ruledtabular}
\caption{Coefficients of the cubic polynomial, $E(q)=E_0 (1 + c_1 q + c_2 q^2/2 + c_3 q^3/6)$, used to fit the dispersion of the onset {\bf F1} along the $\Gamma M$ line, obtained with the different levels of theory and the EELS experiment.
}
\label{tab:gap_mass}
\end{table}

At low energy and momentum, both the experimental and computed onset, {\bf F1}, disperses quadratically with $\mathbf{q}$, as expected from the dispersion of the bands slightly below and above the Fermi level, as shown in Fig.~\ref{fig:ZnO_bs_dos}. We have fitted the full $\Gamma M$ dispersion with a cubic polynomial representing a Taylor expansion truncated at third order, of the form $E(q)=E_0 (1 + c_1 q + c_2 q^2/2 + c_3 q^3/6)$, where $E_0$ is the bandgap. The values of the $c_{1-3}$ coefficients fitted to the theoretical and the experimental spectra are summarized in Table~\ref{tab:gap_mass}, while the polynomials are plotted in Fig.~S4 of the \suppinfo.
We have obtained accurate fits in all the cases ($1- R^2 < 10^{-4}$). 
Notice that the effective mass of the dispersion is obtained from the $c_2$ coefficient, and can be compared to other values in the literature~\cite{Zhang2018PRB}. However, we obtain a finite $c_1$ term for IPA and RPA, thus deviating from a parabolic dependence at low $\mathbf{q}$, while $c_1$ is closer to zero in the experiment and BSE, even if the uncertainty is larger in these two cases. The excitonic effects responsible for the BSE dispersion are discussed in Sec.~\ref{sec:excitons}.

In all the panels, the {\bf F2} peak is fairly non-dispersive in energy while its intensity decreases quickly with $\mathbf q$. After half the $\Gamma M$ distance, small deviations are observed due to the emergence of different states at energies slightly above {\bf F2}. In the case of PBE0, {\bf F2} is separated into two peaks from the beginning. 
In agreement with the experiment, all the theoretical descriptions find that {\bf F3} is a mixture of several peaks, arising from its mixed orbital character discussed in the previous section. However, whereas PBE$+U$ and the experiment exhibit an admixture of dispersive and non-dispersive peaks, they are predominantly non-dispersive for PBE and PBE0. PBE and PBE0 also exhibits a notable valley respectively at around $16.6$~eV and $17.2$~eV, more pronounced for PBE0. This valley is not appreciable in PBE$+U$ nor in the experiment.

 \begin{figure*}
    \centering
    \includegraphics[width=0.98\textwidth]{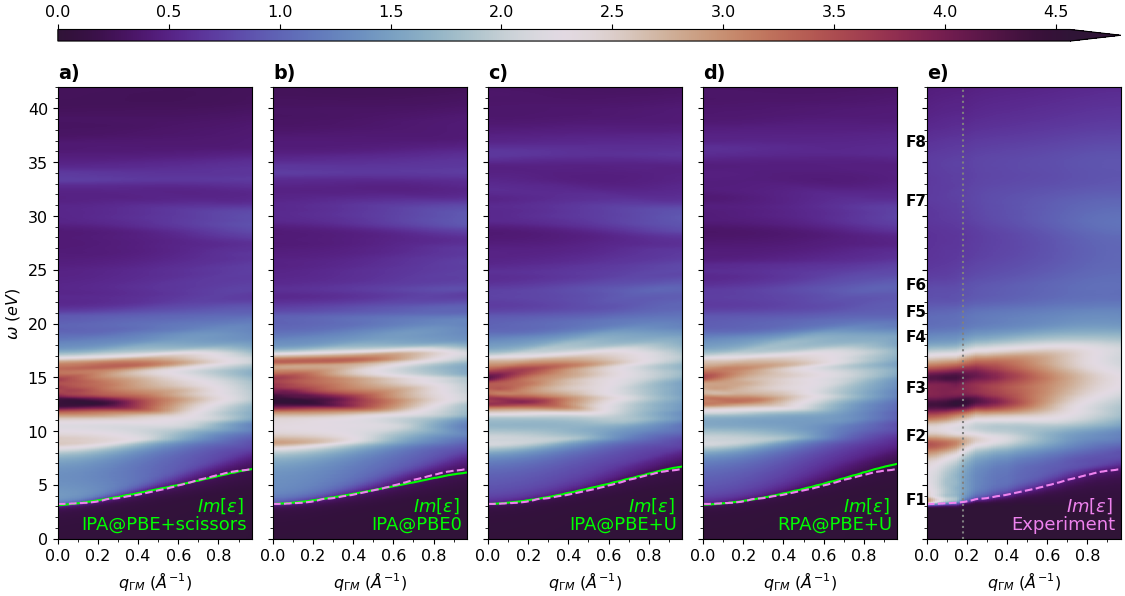}
    \caption{Similar color-maps to Fig.~\ref{fig:cmap_GM_eel} but for the energy and momentum-dependent dielectric function, $\ip[\varepsilon(\mathbf{q},\omega)]$. In the case of the experiment we plot the results of the Kramers-Kronig analysis on the experimental loss function (see App.~\ref{sec:KK_exp}). The same borderlines from Fig.~\ref{fig:cmap_GM_eel} are plotted here to show the similarity of the dispersion of the onset. In panel e) we also plot the same dotted gray line at $q_{\Gamma M}=0.18$~\AA$^{-1}$. In addition, we include labels indicating the position of the {\bf F1-4} features defined for the loss function in Fig.~\ref{fig:ZnO_loss}.}
    \label{fig:cmap_GM_eps}
\end{figure*}

The plasmon ({\bf F4}) and all the following features ({\bf F5-8}), are non-dispersive in their energy position, while changing mainly 
in intensity with $\mathbf{q}$, as found in both theory and experiment. This non-dispersive nature is related to the localization of the $d$-states, which have very flat dispersions in the band structure of Fig.~\ref{fig:ZnO_bs_dos}, and their dipole elements dominating the tail of the spectra. In all the theoretical approaches, the intensity of the double-peaked plasmon {\bf F4} is larger for the peak at higher energy, even if PBE and PBE0 present very different relative intensities with respect to PBE$+U$ results. For $q_{\Gamma M} \leq 0.2$~\AA$^{-1}$, the intensity of the first peak is slightly larger in the experiment, while the intensity of both peaks is very similar at larger $\mathbf{q}$-values, with the second one becoming slightly larger. The decreasing intensities of {\bf F4-6} with increasing $\mathbf{q}$ are related to the quasi-particle nature of the plasmon. The states forming {\bf F7} tend to overlap more at finite momentum, thus increasing the intensity of this feature with $\mathbf{q}$, while {\bf F8} remains almost unaltered. 

In Fig.~\ref{fig:cmap_GM_eps}, we show color-map plots of the momentum-dependent dielectric function, $\ip [\varepsilon(\mathbf{q},\omega)]$, analogous to Fig.~\ref{fig:cmap_GM_eel}, where the experimental spectra were obtained by Kramers-Kronig analysis with the procedure described in App.~\ref{sec:KK_exp}. 
We have found a similar level of agreement between theory and experiment, as for the loss function. These complementary results allow us to better understand the collective nature of the excitations responsible for the features observed in Fig.~\ref{fig:cmap_GM_eel}.

Fig.~\ref{fig:cmap_GM_eps} exhibits many dispersive and non-dispersive features, where PBE$+U$ results show a better agreement with the experimental dispersion, as in Fig.~\ref{fig:cmap_GM_eel}. 
The largest intensities in Fig.~\ref{fig:cmap_GM_eps} are found for the peaks in the region between $8$~eV and $18$~eV, 
since plasmons are less relevant for $\ip[\varepsilon]$ than for $\ip[-\varepsilon^{-1}]$~\cite{martin2016book}. 
There are peaks in $\ip[\varepsilon]$ analogous to the features {\bf F2} and {\bf F3} defined for $\ip[-\varepsilon^{-1}]$, at energies $\sim 0.5$~eV below. {\bf F2} and {\bf F3} are then interpreted as plasmon resonances of the corresponding excitations in $\ip[\varepsilon]$, i.e., the excitations in $\ip[-\varepsilon^{-1}]$ and $\ip[\varepsilon]$ come respectively from long and short range electronic interactions. 
Due to the larger non-dispersive character in PBE and PBE0 results compared to PBE$+U$, there is an effective state superposition that increases the maximum intensity and shows an apparent better agreement of PBE and PBE0 with the experimental intensity. In addition, the maximum intensity decreases at the RPA level with respect to IPA. 
The experiment also shows a large fraction of the spectral weight concentrated around the optical gap. Excitonic contributions, which are not accounted for in the IPA and RPA approximations, are considered at the BSE level in Sec.~\ref{sec:excitons}, providing a better description of the intensity with respect to IPA and RPA.

Fig.~\ref{fig:cmap_GM_eps} also shows several crossings of excitation states corresponding to {\bf F3}, in particular a very intense one around $12.5$~eV, with its vertex around $q_{\Gamma M}\sim 0.25$~\AA$^{-1}$. Some of these crossings in $\ip[\varepsilon]$ can be related to band crossings or reorderings in the band structure of Fig.~\ref{fig:ZnO_bs_dos}. These crossings are also present in the experiment, although not all of them well appreciated in the color-map plot due to limitations in the experimental resolution, especially for $\mathbf{q} \to 0$ and around the the border between the two experimental data sets. 

 \begin{figure*}
    \centering
    \includegraphics[width=0.99\textwidth]{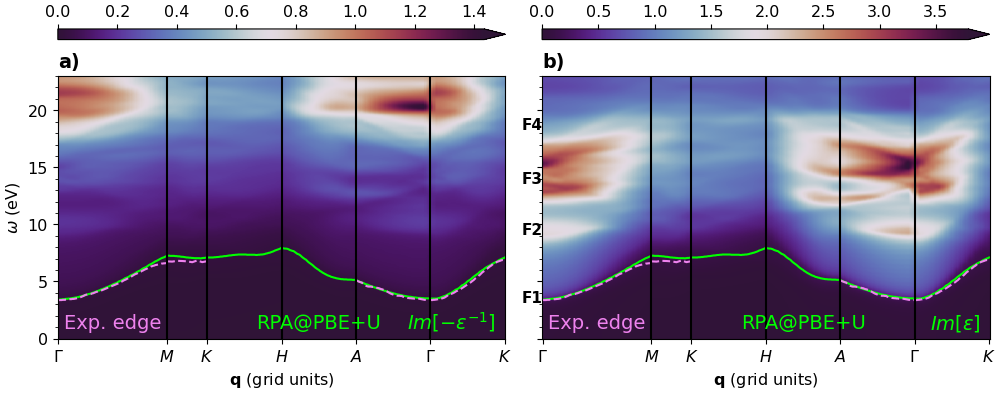}
    \caption{Color-map representation of the energy and momentum-dependent loss function, $\ip[-\varepsilon^{-1}(\mathbf{q},\omega)]$,  (left) and dielectric function,  $\ip[\varepsilon (\mathbf{q},\omega)]$, (right) along several high-symmetry lines in the Brillouin zone, computed at the RPA@PBE$+U$ level. Similar borderlines to Fig.~\ref{fig:cmap_GM_eel} are plotted, including the experimental edge when available. The same labels indicating the position of the experimental {\bf F1-4} features are shown.}
    \label{fig:cmap_bs}
\end{figure*}

 \begin{figure}
    \centering
    \includegraphics[width=0.49\textwidth]{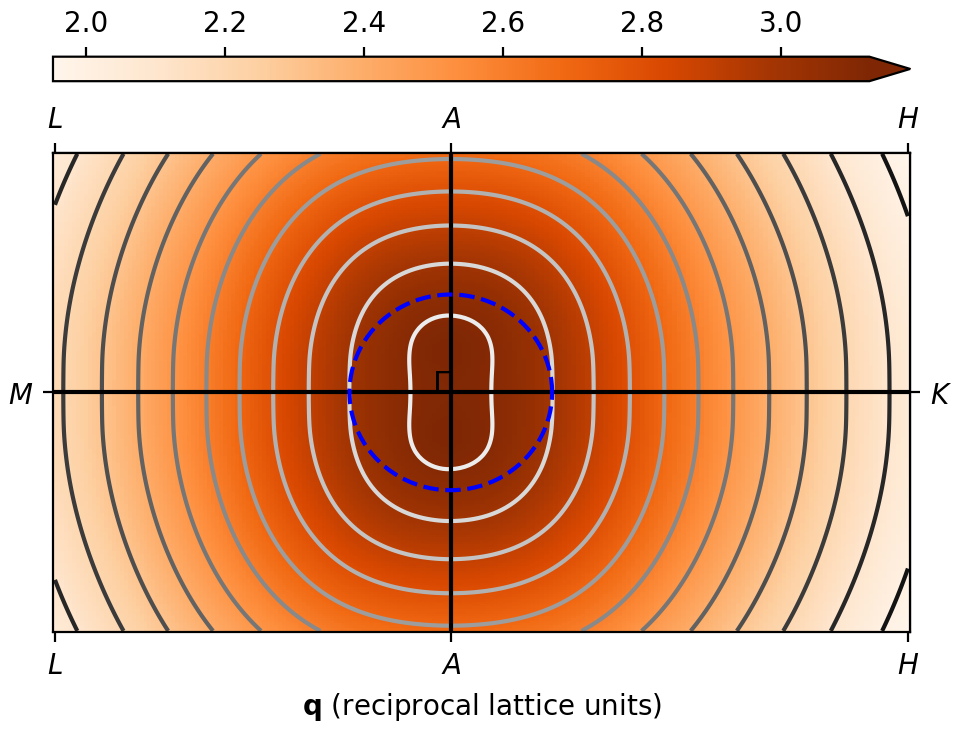}
    \caption{Color-map representation of the momentum-dependent static dielectric constant, $\rp[\varepsilon(\mathbf{q}, \omega=0)]$ of ZnO, computed at the IPA@PBE$+U$ level of theory along in-plane and out-of-plane projections of the Brillouin zone. Contour isolines (gray shades) show anisotropy between the in-plane and out-of-plane directions at finite momentum. The dashed blue circumference, which is tangent to the second smaller isoline in the $M \Gamma K$ lines, represents an ideal isotropic contour.}
    \label{fig:cmap_Xr}
\end{figure}

In Fig.~\ref{fig:cmap_bs} we focus on the region below $23$~eV and extended the range of $\mathbf{q}$-values to several high-symmetry lines in the Brillouin zone, matching the ones used in the band structure plotted in Fig.~\ref{fig:ZnO_bs_dos}, for both $\ip[-\varepsilon^{-1}]$ and $\ip[\varepsilon]$. These excitation band structures correspond to RPA@PBE$+U$ results, while a full comparison of analogous plots at the IPA level on top of PBE, PBE0, and PBE$+U$ is provided in Sec.~III of the \suppinfo. The experimental dispersion of the onset around the subset of measured lines is included in the figure, showing an excellent agreement with the theory. 
We can also track how the other features change along the different directions, as similarly done for the experimental data in Ref.~\cite{Wang2007Micron} along a selected in-plane and out-of-plane directions. The dispersion exhibits diverse and complex behaviors in both panels, showing several states vanishing, emerging, crossing, or splitting, especially for the multiple states in the zone from $8$~eV to $18$~eV. 
However, the most notable property is the strong anisotropy at $\Gamma$ in the out-of-plane $\Gamma A$ line, compared to the in-plane lines $\Gamma M$ and $\Gamma K$, as also found in previous works~\cite{Zhang2006Micron,Wang2007Micron,Gori2010PRB,Zhang2018PRB,Yeh2021PRApp}.

Focusing on $\ip[-\varepsilon^{-1}]$, there are features along $\Gamma A$ not present in the in-plane lines, for example the peak between {\bf F2} and {\bf F3} at $\sim 12$~eV, as also identified in Refs.~\cite{Wang2007Micron,Yeh2021PRApp}. Other states in the zone of {\bf F3} start to appear at finite momentum, reaching their maximum intensity at $\mathbf{q}=A$. The anisotropy manifests even clearer in the region of the plasmon.
Comparing the $\ip[\varepsilon]$ and $\ip[-\varepsilon^{-1}]$ band structures, the superposition of low-range excitations (red and dark orange features in $\ip[\varepsilon]$) and their corresponding plasmon resonances (light blue features in $\ip[-\varepsilon^{-1}]$) is much less overlapped along $H A \Gamma$ than in the $\Gamma M K \Gamma$ lines. In fact, the out-of-plane lines present a very well-defined zigzag sequence of hills and valleys that alternate from $\ip[\varepsilon]$ to $\ip[-\varepsilon^{-1}]$. In other words, the hills and valleys above $8$~eV in $\ip[\varepsilon]$ can be identified respectively as valleys (dark blue features) and hills (light blue features) in $\ip[-\varepsilon^{-1}]$. 
The origin of the zigzag spectra in the anisotropic direction can be related to the band crossings around the middle of the $\Gamma A$ line shown in Fig.~\ref{fig:ZnO_bs_dos}. The anisotropy in the $\Gamma A$ direction is confirmed at the BSE level and the experiment in Sec.~\ref{sec:excitons}.

 \begin{figure*}
    \centering    \includegraphics[width=0.98\textwidth]{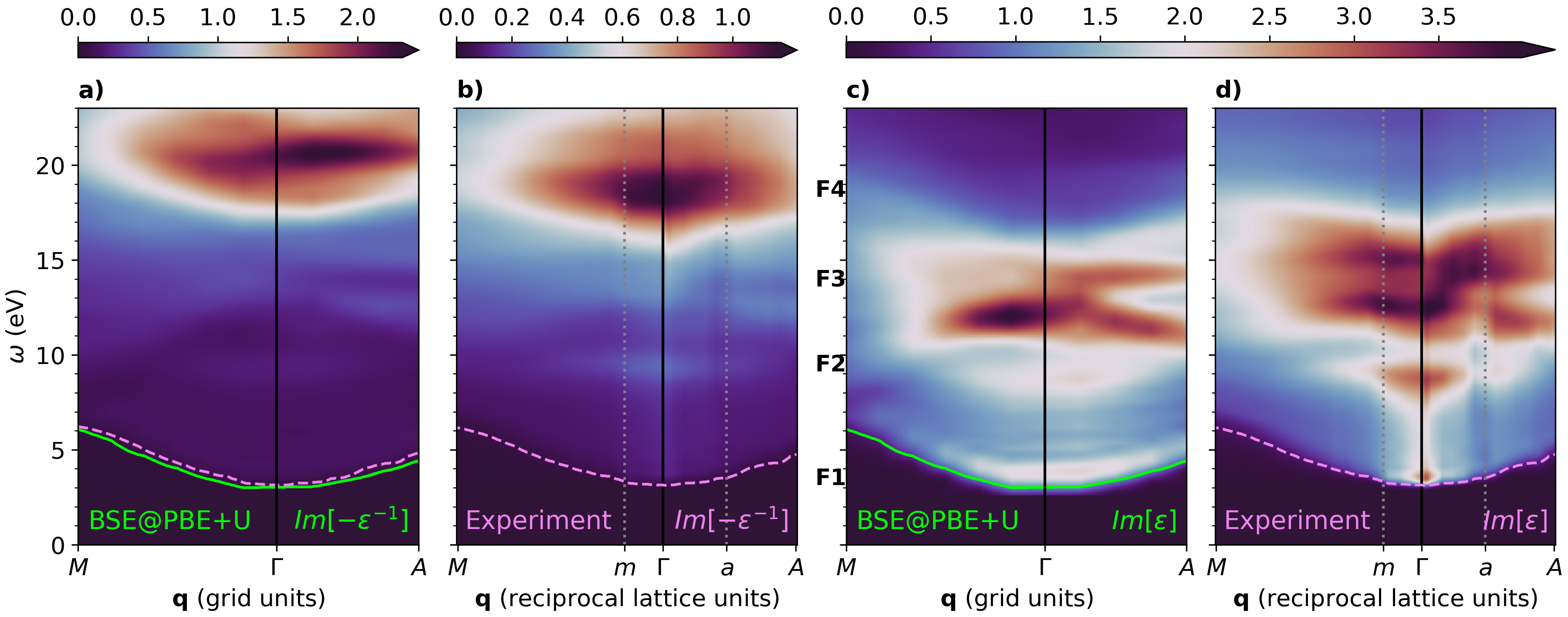}
    \caption{Color-map representation of the energy and momentum-dependent loss function, $\ip[-\varepsilon^{-1}(\omega,\mathbf{q})]$ (panels a-b), and dielectric function,  $\ip[\varepsilon]$ (panels c-d), computed at the BSE@PBE$+U$ level and compared to the experiment along the $M \Gamma A$ lines. Two points beside the high-symmetry ones are marked in the experimental panels, $m$ and $a$, which correspond to $q_{\Gamma M} = 0.18$~\AA$^{-1}$ and $q_{\Gamma A} = 0.30$~\AA$^{-1}$ respectively, the 
    dotted lines at these points separate different experimental data sets (see description in Sec.~\ref{sec:exp_details}).
    Similar borderlines from Fig.~\ref{fig:cmap_GM_eel} are plotted, and the same labels indicating the position of the experimental {\bf F1-4} features. 
    }
    \label{fig:cmap_bse}
\end{figure*}

In Fig.~\ref{fig:cmap_bs} we have included $\mathbf{q} = K$ at the intersection of the in-plane $M K$ and out-of-plane $K H$ lines. At this point of the Brillouin zone border, the anisotropy is less evident than in the case of $\Gamma$ in the $A \Gamma K$ lines, and the limit $\mathbf{q} \to K$ appears continuous in the plot. However, there should be indications of the anisotropy at finite $\mathbf{q}$ as well, since it is a geometric property of the system. In order to address this question at a simple level of approximation, we focus on zero frequency and compute the IPA@PBE$+U$ static dielectric constant, $\rp[\varepsilon(\mathbf{q}, \omega=0)]$, of ZnO at several in-plane and out-of-plane finite $\mathbf{q}$-values. The selected 2D projections of the Brillouin zone are displayed in Fig.~\ref{fig:cmap_Xr}.

As mentioned in the introduction, the static macroscopic dielectric constant $\rp[\varepsilon(\mathbf{q}\to 0, \omega=0)]$ is anisotropic. Although there is a range of experimental values reported in the literature~\cite{Adachi2004handbook,Adachi2013optical}, the difference between the static dielectric constant in the in-plane and out-of-plane directions is around $0.9$. In our IPA@PBE$+U$ calculations shown in Fig.~\ref{fig:cmap_Xr}, the static dielectric constant decreases quadratically from $3.13$ for vanishing $\mathbf{q}$ to $1.95$ at the border of the Brillouin zone. In the $\mathbf{q}\to 0$ limit we find a difference of $0.02$ between the in-plane and out-of-plane directions. The accuracy of these numbers is limited by the incompleteness of the $f$-sum rule~\cite{Egerton2011book,martin2016book} and the level of approximation, since we use a finite number of 100 bands and neglect the local-field effects. In addition, the inclusion of excitonic interactions is expected to increase the intensity of the onset {\bf F1}, and therefore affect the dielectric constant at small frequencies. 
Even if we get a much lower difference with respect to the experimental references, we can still analyze the anisotropy at finite $\mathbf{q}$.

In Fig.~\ref{fig:cmap_Xr} we include contours at fixed values of the dielectric constant. The shape of these isolines (gray shades) is asymmetric with respect to $\pi/2$ rotations due to the anisotropy between in-plane and out-of-plane directions, as compared with an ideal isotropic contour (dashed blue circumference). The radius of the contours along the $\Gamma M$ and $\Gamma K$ is the same, as the response is isotropic along these lines, while the radius along $\Gamma A$ is larger. Since the overall intensity decreases quadratically with $\mathbf{q}$, a larger $\Gamma A$ radius compensates for a larger dielectric constant along this line. Moreover, the difference between the in-plane and out-of-plane radii remains constant at $\Delta q \approx 0.08$~\AA$^{-1}$, for contours sufficiently far from $\Gamma$, resulting in parallel ellipsoidal contours. Due to the decay of the intensity with $\mathbf{q}$, the difference between the dielectric constant in the in-plane and out-of-plane directions also decays quadratically, and therefore the anisotropy is less appreciable at the border of the Brillouin zone.

\begin{figure*}
    \centering
    \includegraphics[width=0.96\textwidth]{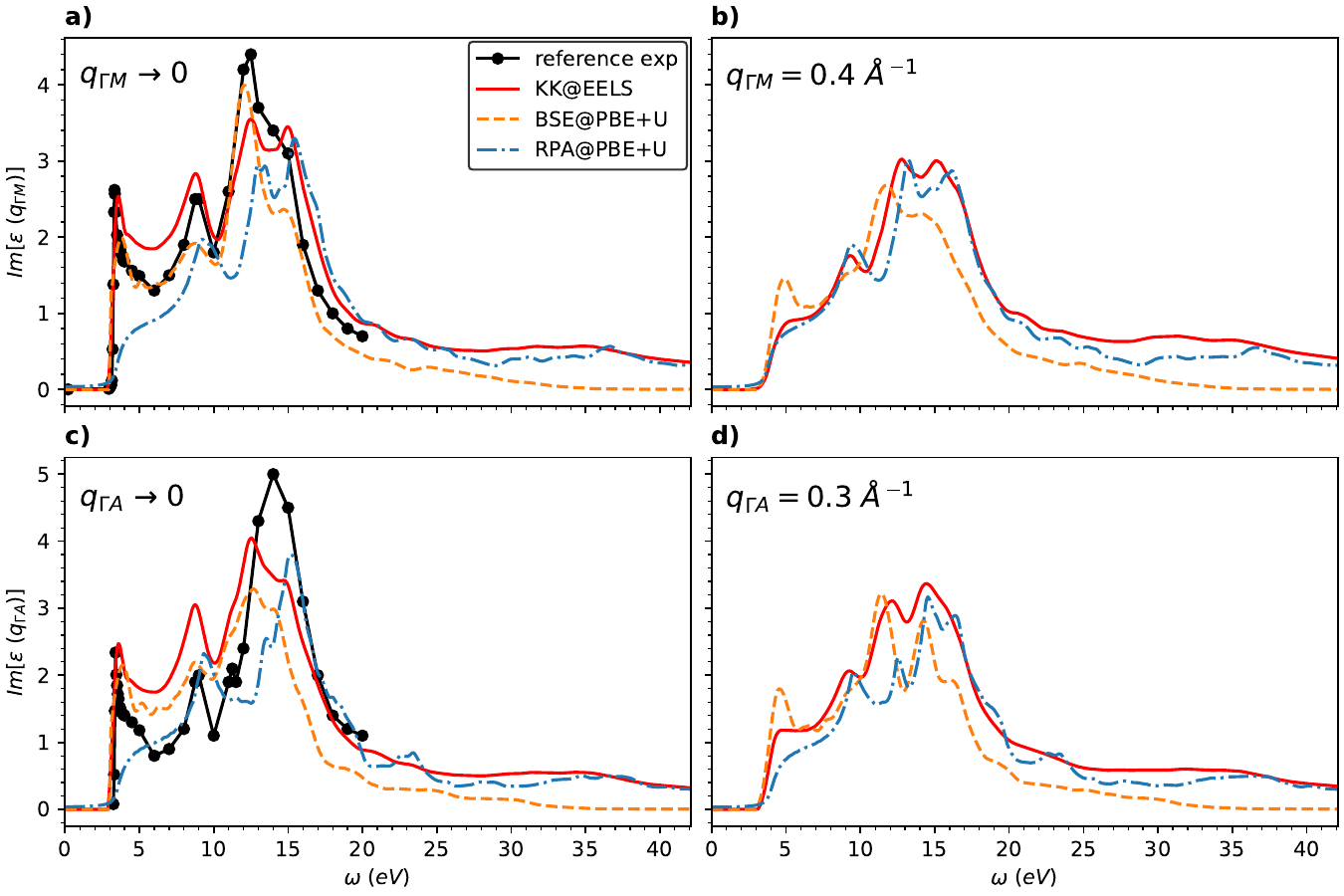}
    \caption{ Comparison of the experimental and theoretical absorption spectra, $\ip[\varepsilon (\mathbf{q},\omega)]$, in the $\Gamma M$ (a, b) and $\Gamma A$ (c, d) lines in the $\mathbf{q}\to 0$ limit (a, c) and for finite momentum (b, d), $q_{\Gamma M}\approx 0.4$~\AA$^{-1}$ and $q_{\Gamma A}\approx 0.3$~\AA$^{-1}$. Black circles correspond to the reference experimental data extracted from Tables D2-1 and D2-2 of Ref.~\cite{Adachi2013optical}, solid red lines correspond to the dielectric function obtained by Kramers-Kronig analysis on the EELS data (see App.~\ref{sec:KK_exp}), and dashed orange and dashed-dotted blue lines correspond respectively to BSE and RPA calculations on top of DFT$+U$.}
    \label{fig:eps_bse}
\end{figure*}

\subsection{Excitonic effects}
\label{sec:excitons}
As described in Sec.~\ref{sec:response}, excitonic effects in ZnO cause a very sharp step-like onset in the EELS spectrum in the optical limit. These effects are highlighted in the imaginary part of the dielectric function, obtained by Kramers-Kronig analysis. In this case, the onset becomes a prominent peak, characteristic of an excitonic nature resulting in a lower optical gap, as found in other experiments measuring directly the absorption spectrum~\cite{Adachi2013optical,RakelIBOOK2008}. 
The fine structure of the excitonic contributions is very rich, presenting several free and bound excitons with different temperature-dependent behaviors~\cite{Teke2004PRB}. Additional intrinsic and extrinsic fine structures such as polaritons, two-electron satellites, donor-acceptor pair transitions, and longitudinal optical-phonon replicas have also been observed in time-resolved photo-luminescence experiments~\cite{Teke2004PRB}.

By describing intrinsic excitonic effects at the BSE level of theory, we here focus on the main broad effects on the computed spectra, since our EELS measurements do not resolve the fine excitonic structures. Moreover, due to the high computational cost of the BSE calculations, the accuracy of the resulting spectra is limited by the convergence parameters specified in Sec.~\ref{sec:com_dets}. Despite this limitation, the BSE calculations provide key insights to understand the role played by electron-hole interactions in the response functions of ZnO.

The excitonic binding energy between the top of the valence and the bottom of the conduction band is experimentally found to be around $60$~meV~\cite{Thomas1960JPCS,Jellison1998PRB,Muth1999JAP,Teke2004PRB}. Our calculations with the $18\times 18\times 12$ $\mathbf{k}$-grid result in a four times larger value, as the binding energy convergences very slowly with the number of $\mathbf{k}$-points. Nonetheless, our values are comparable to the ones found in Ref.~\cite{Zhang2018PRB} with similar grids. Furthermore, extrapolating to grids finer than $36\times36\times24$ results in a binding energy within $\sim 5$~meV of the experimental value, as shown in Ref.~\cite{Zhang2018PRB}.

In Fig.~\ref{fig:cmap_bse}, we present color-map plots of the $\mathbf{q}$ dispersion of the inverse (panels a-b) and direct dielectric functions (panels c-d) in the same energy range of Fig.~\ref{fig:cmap_bs}. In this case, we compare BSE@PBE$+U$ results with the experiment along the $M \Gamma A$ lines. 
We have applied a stretching correction to compensate for the larger excitonic binding energy of our BSE calculations, and thus align the onset with the experimental.
As in the previous IPA and RPA calculations, at the BSE level of theory we can recognize 
the features discussed in Sec.~\ref{sec:q_response}, although they are blurred due to the coarser $\mathbf{k}$-grid used in BSE. 
The poor discretization also causes oscillations on the flat area of the onset below {\bf F2}. Moreover, the limited number of bands provides a maximum excitation energy of $\sim 35$~eV, which is not enough to obtain the features above the plasmon. High-energy features like plasmons are not usually explored with BSE calculations due to the high computational cost. The trends in our unconverged BSE calculations, nonetheless, suggest that electron-hole interactions lower the plasmon energy by $\sim 1$~eV compared to RPA and IPA, resulting in a much better agreement with the experiment. Compared to earlier work, we also note that the BSE loss function of ZnO was computed in Ref.~\cite{Schleife2009PRB} in the optical limit, while here we provide results for finite momentum. 

\begin{figure*}
    \centering
    \includegraphics[width=0.99\textwidth]{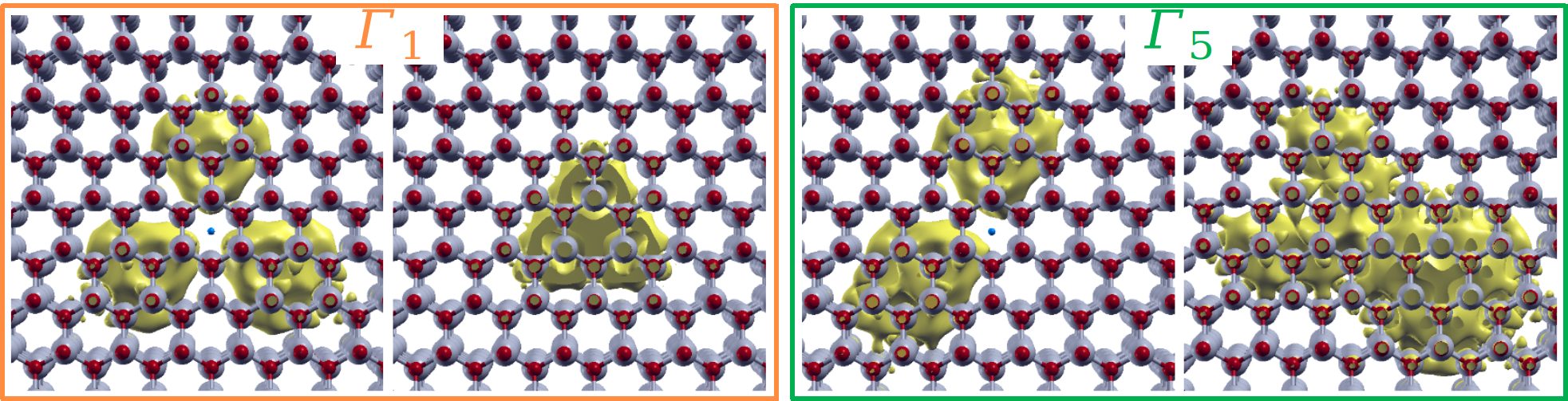}
    
    \caption{Symmetries of the excitonic wave-functions manifested by the A, B and C excitons of ZnO (see Table~\ref{tab:ewf}). The contours (yellow) correspond to the real-space probability distribution of finding a hole when an electron is placed at the dot in the center (blue), and vice-versa. View along the direction perpendicular to the hexagonal plane of the crystalline structure of Zn (gray) and O (red) atoms.}
    \label{fig:exc_wf}
\end{figure*}

At small energies, the dispersion of the onset {\bf F1}, indicated by the solid green and dashed purple lines in 
Fig.~\ref{fig:cmap_bse}, is similar in the theory and the experiment (see also Table~\ref{tab:gap_mass}). However, the dispersion of the intensity of the excitonic peak is qualitatively different. Where the theoretical decreases quadratically following the energy dispersion, the experimental drops abruptly, in comparison. 
As mentioned above, there are accuracy limitations in the theory, due to the difficulties in localizing the excitons with unconverged $\mathbf{k}$-grids. 
There are also accuracy limitations in the experiment, due to the different data sets delimited by the dotted gray lines at $q_{GM}=m$ and $q_{GA}=a$. However, as commented below, we attribute the main difference in the decay of the intensity to physical effects not described in the theory.

Fig.~\ref{fig:eps_bse} provides a comparison of our theoretical and experimental results of $\ip[\varepsilon(\omega,\mathbf{q})]$ in the $\mathbf{q}\to 0$ limit (panels a, c) and at selected finite values of  $\mathbf{q}$ (panels b, d), in the in-plane (panels a, b) and out-of-plane (panels c, d) directions. In panels (a, c) we also include absorption measurements from Ref.~\cite{Adachi2013optical}. 
The absorption spectra obtained from the EELS measurements, with the methods described in Apps.~\ref{sec:loss_exp} and ~\ref{sec:KK_exp}, show an excellent agreement with previous experimental results~\cite{Adachi2013optical,RakelIBOOK2008}, confirming the viability of the proposed procedures. As for IXSS experiments~\cite{Schulke1988PRB,Schulke1989PRB,Schulke1995PRB,Watanabe2006BCSJ,Weissker2010PRB}, these methods provide EELS with a route to resolve the momentum dependence of the direct dielectric function, $\ip[\varepsilon(\mathbf{q},\omega)]$, not easily accessible in other spectroscopies. 
The figure also shows the very good agreement of the resulting BSE on top of PBE$+U$ spectra with both experiments, and they also agree with ellipsometry measurements from Refs.~\cite{Jellison1998PRB,RakelIBOOK2008,Gori2010PRB}. Only some of the peaks beyond $\sim 13$~eV and the tail of the BSE spectra are lower due to the limited number of bands included in the calculations. 

At finite $\mathbf{q}$, the BSE results are also in good agreement with the experiment, except for the slower diminishing of the intensity of the excitonic peak with transferred momentum, as discussed above. Thus, for the selected values of $\mathbf{q}$ along each direction, the experiment is very close to the RPA spectra, which do not account for excitons. 
The rapid decay of the experimental intensity, shown in both Figs.~\ref{fig:cmap_bse} and ~\ref{fig:eps_bse}, is likely related to the different dispersion of the excitonic fine structure of ZnO~\cite{Teke2004PRB},
including exciton-phonon and exciton-defect coupling. 
However, analyzing the effect of such many-body effects~\cite{Teke2004PRB}, is out of the scope of this work.

The BSE results in Figs.~\ref{fig:cmap_bse} and ~\ref{fig:eps_bse} confirm the anisotropy of the features described in Sec.~\ref{sec:q_response} at lower levels of theory. In addition, they show anisotropy in the onset of the direct dielectric function, as found experimentally. We investigate this excitonic anisotropy by analyzing the excitons formed by the coupling of the top three valence states at the $\Gamma$ point with the first conduction state. These excitons are usually denoted as A, B, and C in the literature~\cite{Park1966PR,Liang1968PRL,Reynolds1999PRB,Teke2004PRB,Tsoi2006PRB,Gori2010PRB}.

From theoretical arguments on the symmetry ordering in ZnO (wurtzite crystal)~\cite{Park1966PR,Reynolds1999PRB,Teke2004PRB} and the band structure in Fig.~\ref{fig:ZnO_bs_dos}, the lowest conduction state is predominantly of Zn-$s$ type and has $\Gamma_7$ symmetry, while 
the top three valence states are predominantly of O-$p$ type and have symmetries $\Gamma_9$, upper $\Gamma_7$ and lower $\Gamma_7$ respectively. 
The $\Gamma_9 \times \Gamma_7 \to \Gamma_5 + \Gamma_6$  and $\Gamma_7 \times \Gamma_7 \to \Gamma_5 + \Gamma_1+ \Gamma_2$ combinations generate the excitons~\cite{Reynolds1999PRB,Teke2004PRB}.
The energy splitting of the states due to spin-orbit interactions~\cite{Teke2004PRB} is smaller than the energy resolution of our EELS experiments, therefore we have not considered them in our calculations.
In Fig.~\ref{fig:exc_wf} we show the two types of excitonic wave-functions we obtain for the A, B, and C excitons, which
have $\Gamma_1$ and $\Gamma_5$ symmetries. The composition is however asymmetric in the $\Gamma M$ and $\Gamma A$ lines. As summarized in Table~\ref{tab:ewf}, in the $\Gamma M$ the A and B wave-functions have $\Gamma_1$ symmetry and the C exciton is of type $\Gamma_5$, while in the $\Gamma A$ line A and B are of type $\Gamma_5$ and C of type $\Gamma_1$.

\begin{table}
\centering
\begin{ruledtabular}
\begin{tabular}{lccc}
\\[-3pt]
            &  A         &  B         &  C \\[5pt]
\hline\\[-3pt]

$\Gamma M$  & $\Gamma_1$ & $\Gamma_1$ & $\Gamma_5$             \\
$\Gamma A$  & $\Gamma_5$ & $\Gamma_5$ & $\Gamma_1$             \\
\end{tabular}
\end{ruledtabular}
\caption{Symmetry of the excitonic wave-functions corresponding to the A, B and C excitons of ZnO computed at the $\Gamma$ point in the $\Gamma M$ and $\Gamma A$ lines.
}
\label{tab:ewf}
\end{table}

\section{Conclusions}
\label{sec:conclusions}
In this work, we have thoroughly studied the low-energy response functions of ZnO, by characterizing their main features up to $42$~eV and their dispersion with transferred momentum. To do so, we have combined momentum-resolved EELS experiments and first-principle calculations at different levels of the many-body theory. 
We have identified and described the orbital character of eight main features in the loss function. The quantitative comparison between theory and experiment, of both, the momentum-dependent direct and inverse dielectric functions, 
allows us to link some of the main low-energy signatures, like the onset of the spectra and their dispersion, with the bandgap and other features of the electronic band structure, like crossings and band reorderings, that are usually measured in angle-resolved photoemission and inverse photoemission experiments. 

Our results emphasize the importance of performing an accurate post-processing of the EELS measurements to obtain a quantitative comparison between the theoretical and the experimental loss functions, especially their dispersion in $\mathbf{q}$-space. 
We have used a procedure based on the $f$-sum rule, computed partially with a finite cutoff energy, to impose a momentum-dependent normalization on the intensity of our EELS measurements and be able to extract the experimental loss function from them. 
The proposed method is a key ingredient to perform Kramers-Kronig analysis on our EELS measurements at finite momentum. The obtained absorption spectra have an accuracy comparable to direct measurements using, e.g., absorption spectroscopies and ellipsometry, while accessing the full $\mathbf{q}$ dependence.

Relevant many-body interactions, such as plasmons and excitons, have been unraveled by systematically increasing the level of the theoretical calculations, from DFT at the PBE, PBE0, and PBE$+U$ level to optical calculations at the IPA, RPA, and BSE level. We highlight the strong anisotropy of ZnO along the perpendicular direction with respect to the hexagonal plane of its crystalline structure, illustrated by very clear signatures in its momentum-resolved dielectric response and its excitonic wave-functions. Moreover, we provide a theoretical and experimental description of the dispersion with transferred momentum of the excitonic peak that gives shape to the onset of the imaginary part of the dielectric function.

\section*{Acknowledgments}
%
This work was funded by the Research Council of Norway through the MORTY project (315330), and the NORTEM national infrastructure (197405) where the experiments were performed. Access to high performance computing resources was provided by UNINETT Sigma2 (NN9711K), where the computations were performed.
We acknowledge stimulating discussions with Justin Wells, Simon Cooil, and Håkon Røst. D.A.L. also acknowledges Claudia Cardoso for valuable comments.

\begin{appendix}
\section{Loss function from EELS measurements}\label{sec:loss_exp}
The double differential cross-section obtained in EELS is in practice affected by the finite momentum and energy resolutions of the experiment. As mentioned in Sec.~\ref{sec:methods}, it is difficult to obtain the loss function directly from the measurements. 
The main effect of the momentum and energy convolution is the different scale and dispersion of the intensity of the spectra as a function of momentum, $\mathbf{q}$. We address this issue by proposing a $\mathbf{q}$-dependent normalization based on the physical constraint given by the $f$-sum rule~\cite{Egerton2011book,martin2016book}.

\begin{figure*}
 \centering
 \includegraphics[width=0.9\textwidth]{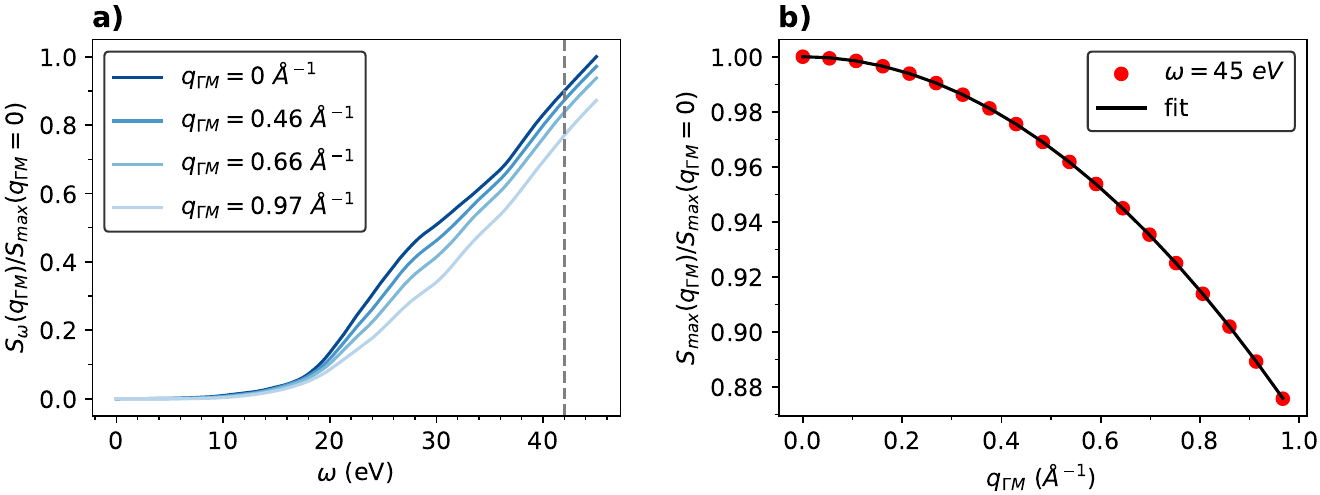}
    \caption{Partial sum rule relative to its value for $q_{\Gamma M}=0$ at a maximum frequency of $45$~eV, corresponding to RPA@PBE$+U$ results. (a) frequency dependence for four values of momentum, $q_{\Gamma M}$. (b) Momentum dependence in the ${\Gamma M}$ interval at the maximum frequency, fitted to a parabola with the form $S(q_{\Gamma M})/S({q_{\Gamma M}=0}) = 1 - a q_{\Gamma M}^2$, where $a = 0.13$~\AA$^2$.}
    \label{fig:sum_rule}
\end{figure*}

As done in Ref.~\cite{Hengehold1970PRB} for the dielectric function, $\ip[\varepsilon]$, we define the following integral for the loss function:
\begin{equation}
    S_{\omega} (\mathbf{q}) = \frac{2}{\pi} \int_0^{\omega} d\omega' \omega' L(\mathbf{q}, \omega').
    \label{eq:sum_rule}
\end{equation}
According to the $f$-sum rule, if the integration is performed to infinity, the resulting quantity, $S_{\infty}$, is independent of $\mathbf{q}$ and proportional to the electronic density of the system. 
This condition has been used in momentum-resolved IXSS experiments to normalize the measured intensities~\cite{Schulke1988PRB,Schulke1989PRB,Schulke1995PRB,Watanabe2006BCSJ,Weissker2010PRB}. In the case of our EELS measurements, the sensitivity of such experiments to multiple scattering processes, more common at larger energies, and the quick decay of the intensity due to the $1/q^2$ factor in Eq.~\eqref{eq:cross_sec}, may introduce undesired variations in the $f$-sum rule~\cite{Weissker2010PRB}, especially for large $\mathbf{q}$. We therefore apply a more strict normalization on our truncated spectra at $45$~eV. 
A partial integration of Eq.~\eqref{eq:sum_rule} accounts for the effective density forming the features at energies below the cutoff value~\cite{Hengehold1970PRB}, which may then induce a $\mathbf{q}$ dependence due to the electronic contributions left out of the integration.

In Fig.~\ref{fig:sum_rule} (panel a) we plot the partial sum rule, $S_{\omega} (\mathbf{q})$, of ZnO numerically integrated with a piecewise linear quadrature rule, as a function of the frequency for four values of transferred momentum along the $\Gamma M$ line. The curves are obtained from RPA@PBE$+U$ results truncated at a maximum energy of $\omega_{\text{max}}=45$~eV. In panel b) we show the momentum dependence of $S_{\omega}$ at a fixed value of $\omega = \omega_{\text{max}}$ in the whole $\Gamma M$ interval. The data is accurately fitted with a parabola, $S/S_{\text{max}} = 1 - a q^2$, where we obtain a value of $a = 0.13$~\AA$^2$ and a $p$-value of the order of $10^{-44}$. In Fig.~\ref{fig:sum_rule} (panel a) we also show a dashed vertical line at $42$~eV, from where the dependence is approximately linear, so that a fit at any fixed frequency in this $3$~eV interval preserves the value of the amplitude of the parabola, $a$, and the quality of the fit. Moreover, at variance with a full integration, a partial integration of the $f$-sum rule is sensitive to the local variations of the electronic density obtained with different levels of theory, however, with the selected cutoff we do not find significant differences in the parameters fitted to PBE, PBE0 and PBE$+U$ results.

If we apply Eq.~\eqref{eq:sum_rule} with the same maximum energy to the EELS measurements, we find a different $\mathbf{q}$ dependence for $S_{\text{EELS}}$, since the intensity of the experiment corresponds to a different quantity. However, we can use two inputs from our RPA calculations to define a $\mathbf{q}$ dependent normalization and find the experimental loss function as
\begin{equation}
    L_{\text{EELS}}(\mathbf{q},\omega) = I_{\text{EELS}}(\mathbf{q},\omega)\frac{S_{\text{RPA}}(1 - a_{\text{RPA}} \mathbf{q}^2)}{S_{\text{EELS}}(\mathbf{q})},
\end{equation}
where $I_{\text{EELS}}$ is the measured intensity, $S_{\text{RPA}}$ is the theoretical sum rule in the $\mathbf{q}\to 0$ limit and $a_{\text{RPA}}$ is the fitted amplitude of the parabolic dispersion. A plot of the resulting experimental loss function of ZnO is provided in Sec.~IV of the \suppinfo. 
The proposed normalization uses an additional parameter, $a$, with respect to previous methods based on the $f$-sum rule, as applied in IXSS~\cite{Schulke1988PRB,Schulke1989PRB,Schulke1995PRB,Watanabe2006BCSJ,Weissker2010PRB}. Note that while here we have computed this parameter at the RPA level, it could be also obtained in other ways, theoretically or experimentally.

\section{Kramers-Kronig analysis}\label{sec:KK_exp}
The real and imaginary parts of the polarizability function, $\chi$, defined in Sec.~\ref{sec:theory}, are related by the Kramers-Kronig relations~\cite{Egerton2011book,martin2016book}. Therefore, we can obtain the complex inverse dielectric function from its imaginary part, by the following (time-ordered) Hilbert transform:
\begin{multline}
     \varepsilon^{-1}(\omega + i \delta)= 1 + \frac{i}{2\pi} \int_{0}^{\infty} 
     d\omega' \ip[\varepsilon^{-1}(\omega')] \times \\ 
     \left[ \frac{1}{\omega-\omega'+ i \delta} - \frac{1}{\omega+\omega'+ i \delta}\right] ,
     \label{eq:Hilbert}
\end{multline}
where $\delta \to 0^{+}$.
We can use the previous equation to perform Kramers-Kronig analysis~\cite{Egerton2011book} on the experimental data and obtain the direct dielectric function $\varepsilon (\omega)$, as also done for the EELS data on ZnO in Refs.~\cite{Zhang2006Micron, Huang2011PRB}. In our case, we apply it to the momentum-resolved experimental loss function, $L(\mathbf{q},\omega)$, obtained with the procedure described in Sec.~\ref{sec:loss_exp}. 
Similar Kramers-Kronig analyses at finite momentum have been performed on IXSS measurements for other materials~\cite{Schulke1988PRB,Schulke1989PRB,Schulke1995PRB,Watanabe2006BCSJ,Weissker2010PRB}.

As for the $f$-sum rule of Eq.~\eqref{eq:sum_rule}, the integration in the Hilbert transform of Eq.~\eqref{eq:dyson} is truncated at the same finite frequency, $\omega'=\omega_{\text{max}}$, and is solved numerically with the same piecewise linear rule. In this case, the partial integration only affects frequencies close to the cutoff, due to the inverse dependence with the integration frequency $\omega'$, as also found in Ref.~\cite{Weissker2010PRB}. To avoid this issue, it is sufficient to extend the integration domain by extrapolating the spectra a few electron-volts beyond the cutoff. The only remaining contribution is the integration of the tail $\omega' \in (\omega_{\text{max}}, \infty)$, which is common to all the frequencies $\omega < \omega_{\text{max}}$ and results in a constant shift of $\rp[\varepsilon^{-1}(\omega)]$~\cite{Egerton2011book}. However, getting a correct static limit $\rp[\varepsilon^{-1}(\omega = 0)]$ is important when inverting $\varepsilon^{-1}$ to get an accurate dielectric function, $\varepsilon$, as also found in Ref.~\cite{Weissker2010PRB}. While in principle we can use other experimental references for the static dielectric constant~\cite{Adachi2013optical}, $\rp[\varepsilon(\omega = 0)]$, we opted to use the completeness of the $f$-sum rule in Eq.~\eqref{eq:sum_rule} to set the weight of the tail in the Hilbert transform.

\end{appendix}

\bibliography{main.bbl}

\end{document}



\title{
Unraveling many-body effects in ZnO: Combined study using momentum-resolved electron-energy loss spectroscopy and first-principles calculations: Supplemental material
 }

\author{Dario A. Leon}%
\email{dario.alejandro.leon.valido@nmbu.no}
\affiliation{
 Department of Mechanical Engineering and Technology Management, \\ Norwegian University of Life Sciences, NO-1432 Ås, Norway
}%

\author{Cana Elgvin}
\affiliation{Centre for Materials Science and Nanotechnology, Department of Physics, University of Oslo, NO-0316 Oslo, Norway
}%

\author{Phuong Dan Nguyen}
\affiliation{Centre for Materials Science and Nanotechnology, Department of Physics, University of Oslo, NO-0316 Oslo, Norway
}%

\author{\O ystein Prytz}
\affiliation{Centre for Materials Science and Nanotechnology, Department of Physics, University of Oslo, NO-0316 Oslo, Norway
}%

\author{Fredrik S. Hage}
\affiliation{Centre for Materials Science and Nanotechnology, Department of Physics, University of Oslo, NO-0316 Oslo, Norway
}%

\author{Kristian Berland}
\affiliation{
 Department of Mechanical Engineering and Technology Management, \\ Norwegian University of Life Sciences, NO-1432 Ås, Norway
}

\date{\today}

\maketitle

\section{Quasi-particle self-consistent $GW_0$ and $GW$}
\begin{figure*}[h]
    \centering  \includegraphics[width=0.69\textwidth]{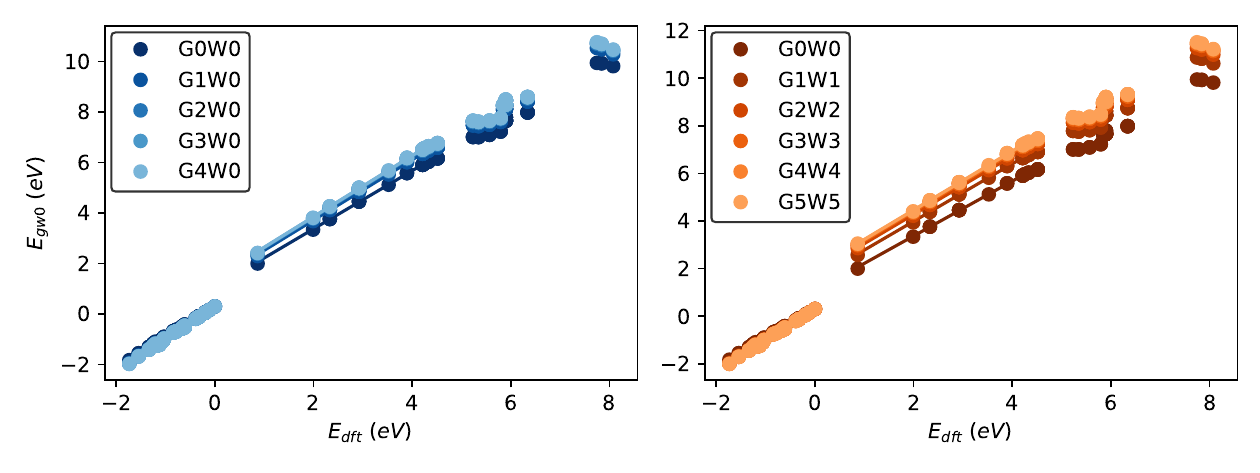}
    \caption{Quasi-particle self-consistent $GW_0$ and $GW$ results starting from DFT@PBE for a set of quasi-particle energies around the fundamental gap of ZnO. The indexes in the legend correspond to the number of iterations, Solid lines in the plot correspond to a linear fit of quasi-particles in the conduction bands.}
    \label{fig:GWself}
\end{figure*}

In Fig.~\ref{fig:GWself} we show eigenvalue self-consistent $GW_0$ and $GW$ data for a set of quasi-particles of ZnO in an energy range from $-2$~eV to $8$~eV. The starting point of the calculations are DFT results obtained with the PBE functional. As expected, both levels of self-consistency increase the gap and stretch the valence and conduction bands. As illustrated in the figure, the $GW$ effects on the PBE band structure can be approximated by a linear transformation using a simple scissors plus stretching operator.
However, without full self-consistency, the hybridization of the states remains that of DFT.



\section{Analysis of dipoles matrix elements}
 \begin{figure*}
 \centering
 \includegraphics[width=0.99\textwidth]{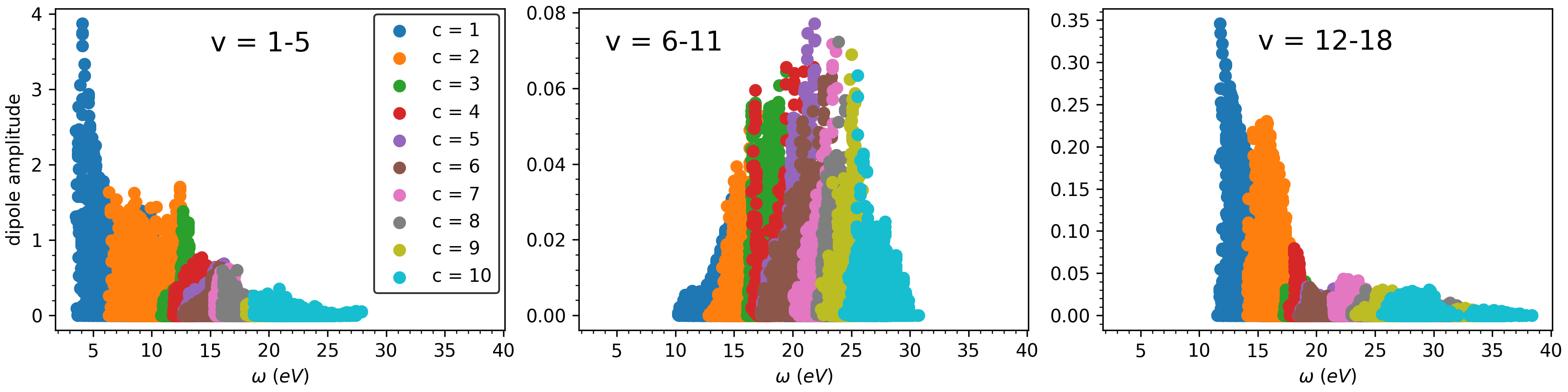}
    \caption{Amplitude of dipole matrix elements of ZnO as a function of their transition energy. Indexing the valence and conduction bands respectively from below and above the Fermi level, the dipoles correspond to transitions from three sets of valence bands: 1-5 (left), 6-11 (center) and 12-18 (right), to the first 10 conduction bands, as indicated in the legend.}
    \label{fig:dipoles}
\end{figure*}
In Fig.~\ref{fig:dipoles} we plot the amplitude of dipoles matrix elements of ZnO  computed from bands around the Fermi level, within the IPA@PBE+U approximation. Each of the three panels corresponds to a different set of valence bands separated in energy (see the band structure in Fig.~1 of the main text). They are subdivided in a color code according to the specific conduction band forming the dipole. We can see that in all the panels each conduction band defines an energy interval for the position of the dipoles, not completely overlapped with the others. Moreover, the top valence bands in the left panel span the larger transition probabilities, which are proportional to their corresponding dipole amplitude. They are followed by the bands of predominantly $d$-character in the right panel, while the valence bands in the intermediate range in the central panel have the lowest transition probabilities. These results are expected from the projected density of states (see the PDOS plot in Fig.~1 of the main text).
%
Due to the small energy range of the onset, it is easy to see, for instance, that the {\bf F1} feature comes from excitations of valence states very close to the Fermi level to the lowest conduction band. 

 \begin{figure*}
 \centering
 \includegraphics[width=0.8\textwidth]{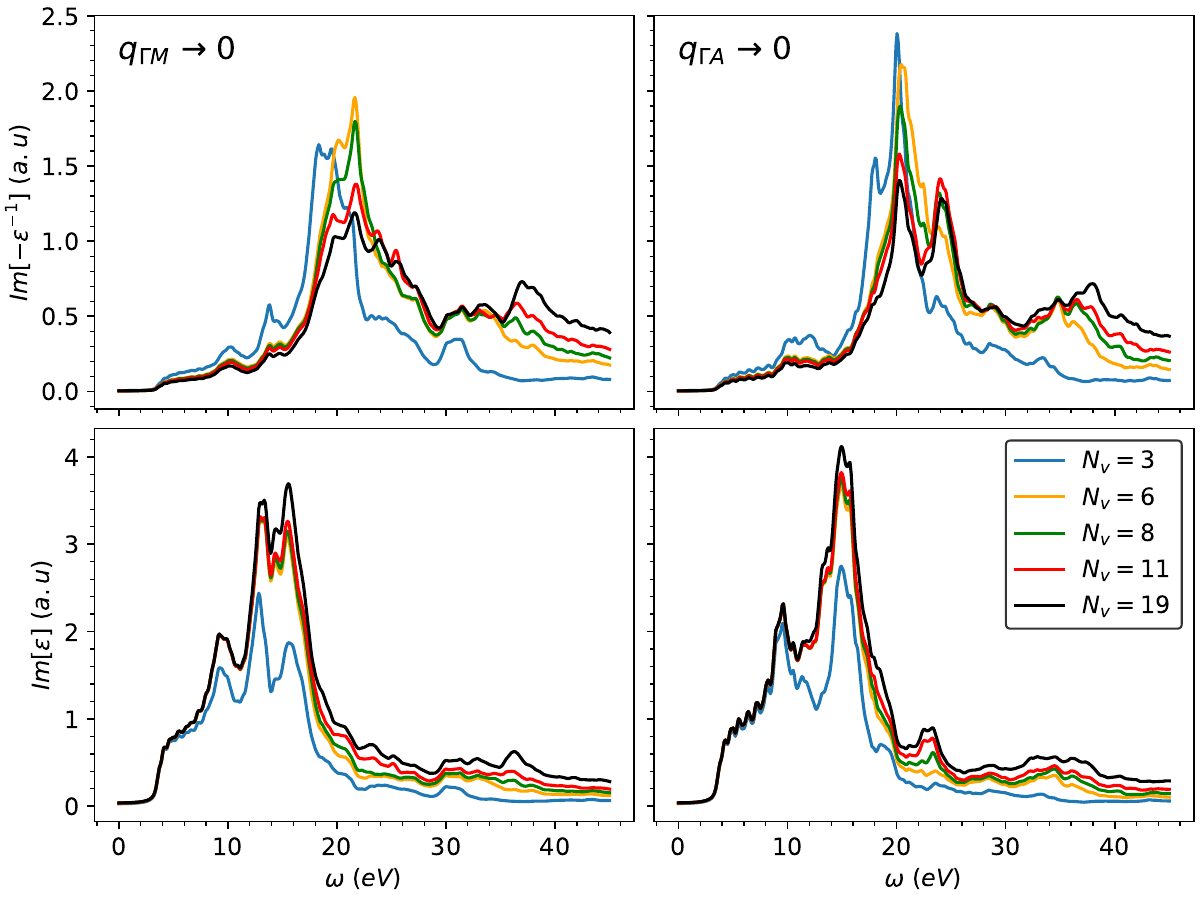}
    \caption{Loss function, $\ip[-\varepsilon^{-1}]$,  (top panels) and the dielectric function,  $\ip[\varepsilon]$, (bottom panels) in the $\mathbf{q}\to 0$ along the $\Gamma M$ (left) and $\Gamma A$ (right) directions. The spectra are computed at the IPA@PBE+U level with a $18\times 18\times 12$ grid of $\mathbf{k}$-points and a reduced number of valence bands specified in the legend. The number of conduction bands is fixed at 74. The case with $N_v=19$ contains all the $d$-states of ZnO and is very close to the full results including the semi-cores states present in the pseudo-potentials.}
    \label{fig:Xip}
\end{figure*}


In order to analyze optical features at larger energies, we have computed the response function of ZnO with a variable number of valence bands. In Fig.~\ref{fig:Xip} we show the results for the direct and inverse dielectric functions in the optical limit. Many of the features appear already with the top three valence bands, while others have a main contribution from states at lower energies, with predominant $d$-character.
Besides {\bf F1}, {\bf F2} is also accurate with respect to the full loss function. Therefore, both features correspond to excitations from O-$p$ states. In the case of {\bf F3}, it shows a well-defined double peak rather than been a shoulder. 
Few more valence states are needed to reproduce this shape. The extra states carries a contribution from Zn-$d$ orbitals, giving a mixed character to {\bf F3}. 

{\bf F4-8} have also mixed characters, 
although not in the same proportion. In particular, the double-peak trait of the plasmon can be found already with O-$p$ contributions, while  we find {\bf F5} and {\bf F6} to have a dominant Zn-$d$ component. Due to the strong dependence on the level of orbital hybridization, these features look very different when comparing PBE and PBE0 with PBE+U (see Fig. 2 in the paper). Moreover, the initial part of {\bf F7} mainly comes from the top three valence bands, while {\bf F8} only has contributions from deeper states. 

\section{Momentum dependent spectra at different levels of theory}
 \begin{figure*}
    \centering
    \includegraphics[width=0.98\textwidth]{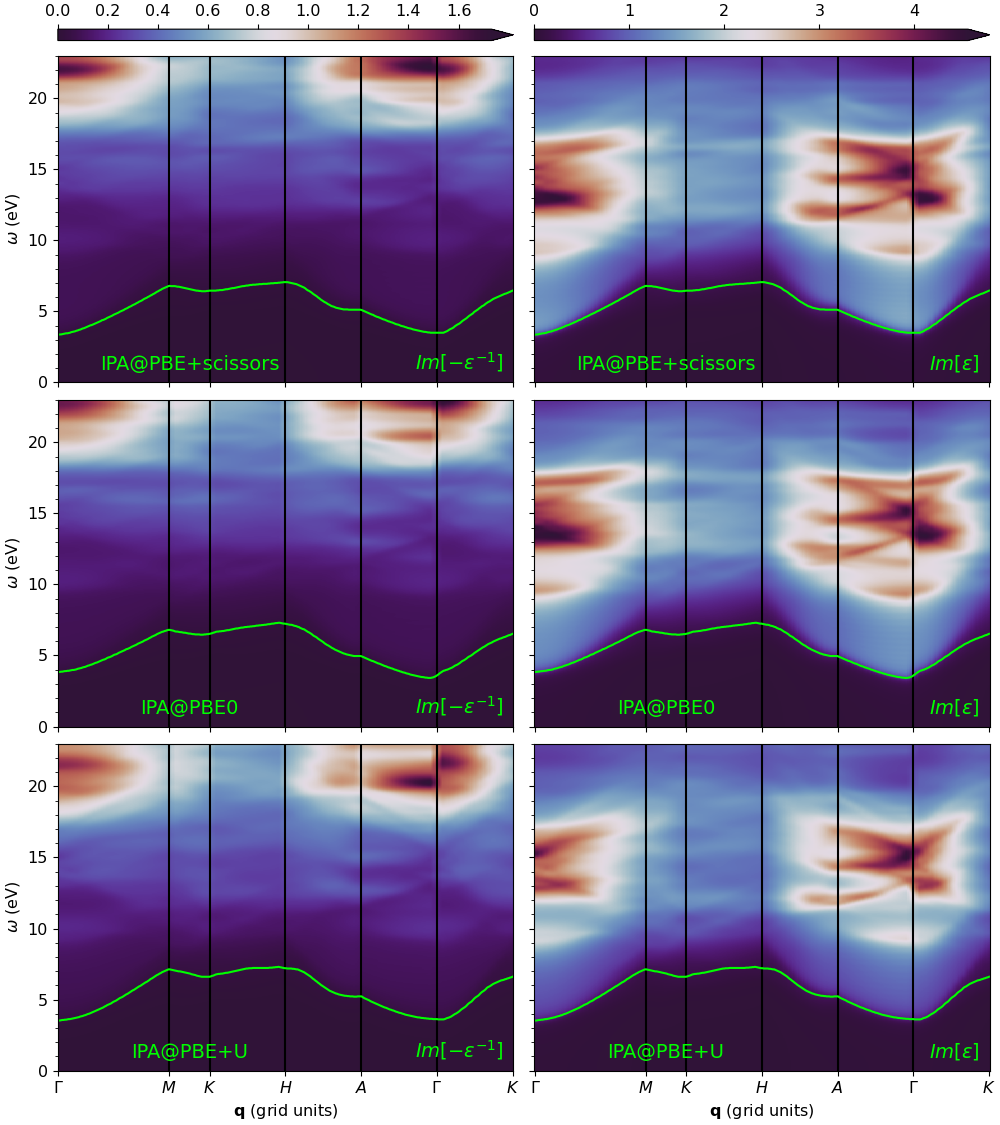}
    \caption{Color map representation of the energy and momentum-resolved loss function, $\ip[-\varepsilon^{-1}(\mathbf{q},\omega)]$,  (left panels) and dielectric function,  $\ip[\varepsilon(\mathbf{q},\omega)]$, (right panels) along several directions in the Brillouin zone, computed at the IPA level on top of PBE (top panels), PBE0 (central panels), and PBE+U (bottom panels). The borderlines highlight the dispersion of the onset of the $\ip[-\varepsilon^{-1}(\mathbf{q},\omega)]$ spectra, and the similarity with the $\ip[\varepsilon(\mathbf{q},\omega)]$ onset.}
    \label{fig:cmap_bs}
\end{figure*}
 \begin{figure}
    \centering
    \includegraphics[width=0.49\textwidth]{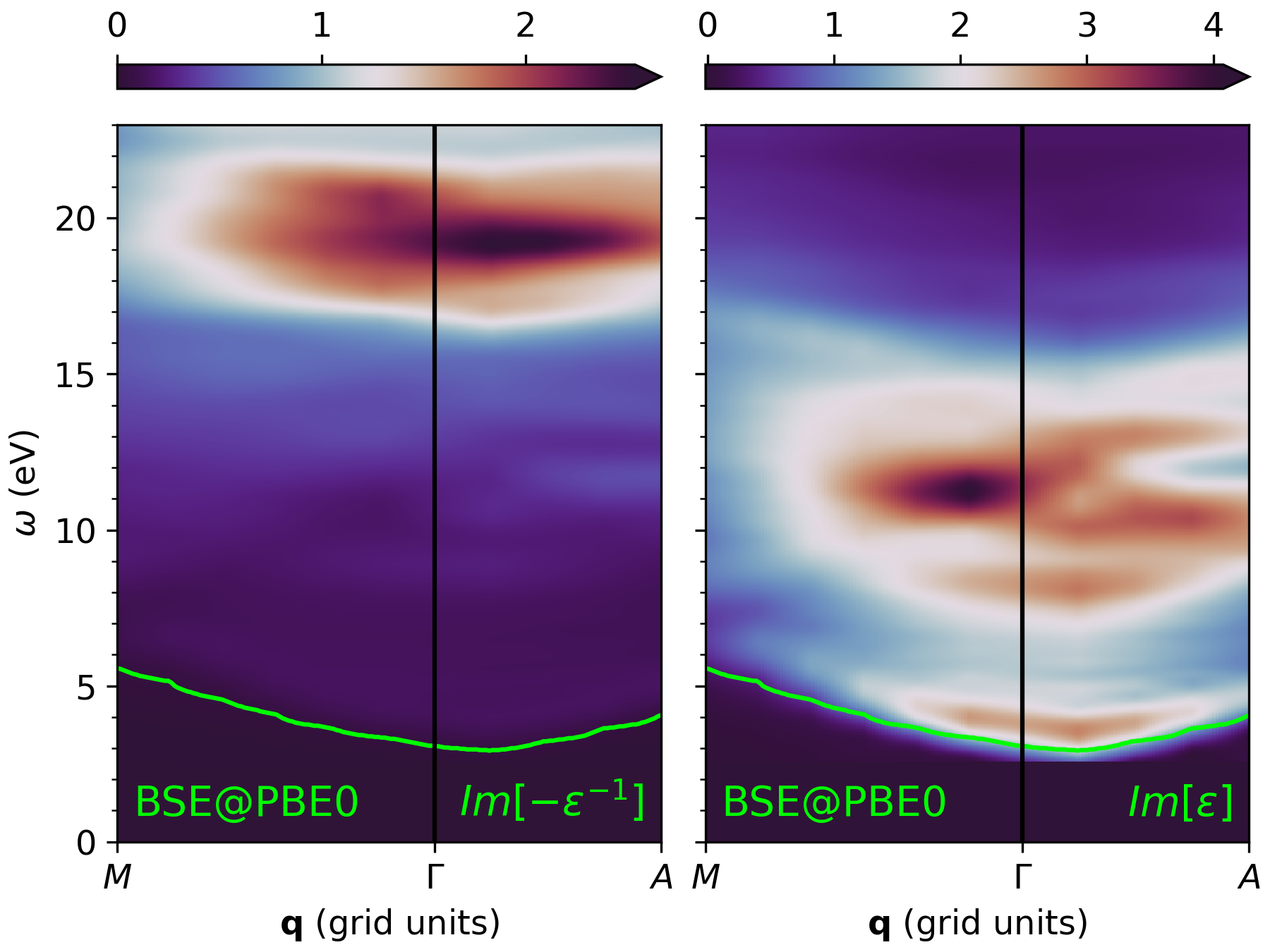}
    \caption{Similar color map plots and borderlines from Fig.~\ref{fig:cmap_bs} for $\ip[-\varepsilon^{-1} (\mathbf{q},\omega)]$ (left) and $\ip[\varepsilon (\mathbf{q},\omega)]$ (right), computed at the BSE@PBE0 level along the $M\Gamma A$ directions.}
    \label{fig:cmap_bse}
\end{figure}
 \begin{figure}
    \centering
    \includegraphics[width=0.49\textwidth]{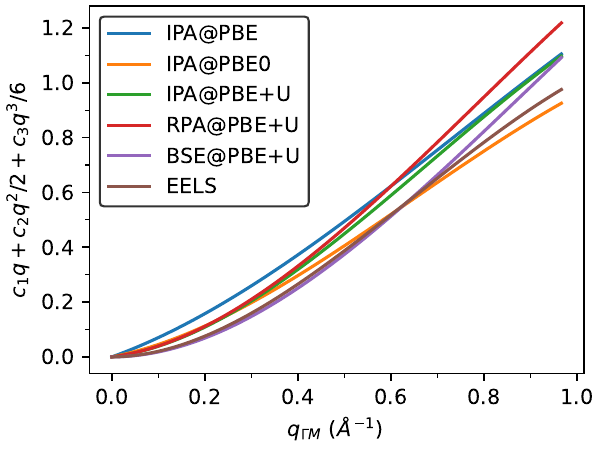}
    \caption{Plot of the cubic polynomial, $E(q)=E_0 (1 + c_1 q + c_2 q^2/2 + c_3 q^3/6)$, used to fit the dispersion of the onset {\bf F1} along the $\Gamma M$ direction, obtained with the different levels of theory and the EELS experiment.}
    \label{fig:f1}
\end{figure}
In Fig.~\ref{fig:cmap_bs} we show color map plots of the momentum-dependent direct and inverse dielectric functions of ZnO, computed with IPA on top of PBE, PBE0 and PBE+U, analogous to the ones in Fig.~5 of the main text for RPA@PBE+U. In Fig.~\ref{fig:cmap_bse} we show similar plots for BSE@PBE0 calculations in the $M \Gamma A$ directions but with a coarser Brillouin zone sampling, corresponding to a $12\times 12\times 8$ grid of $\mathbf{k}$-points. 
In Fig.~\ref{fig:f1} we compare the fitting of the dispersion of the onset along the $\Gamma M$ direction, obtained with the different theories and the experiment.

\section{Post-processing of the experimental data}
\begin{figure*}
    \centering  
    \includegraphics[width=0.98\textwidth]{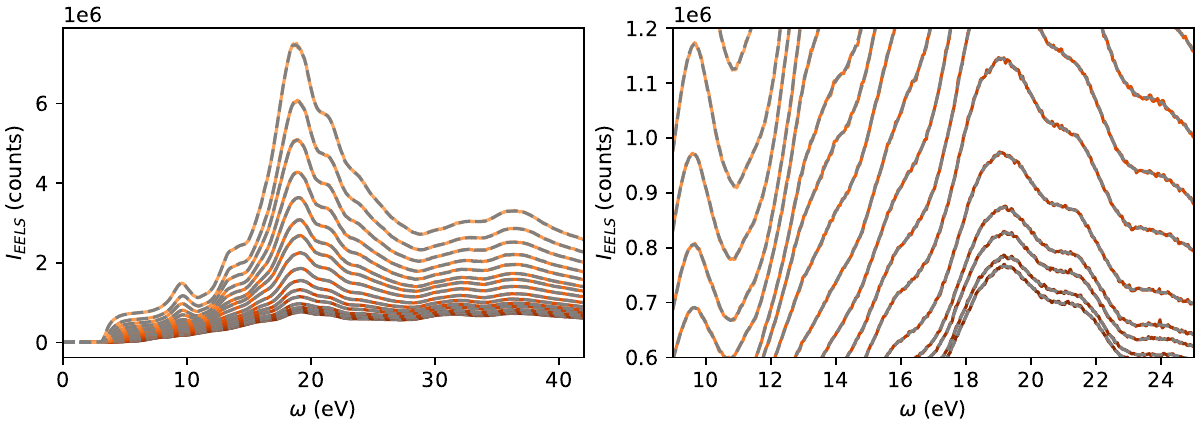}
    \caption{Raw intensity in electron counts measured in momentum-resolved EElS experiments as a function of the energy-loss, along the $Mm$ direction defined in the main text, before (orange shades) and after (dashed gray) the Savitzky-Golay filter (see details in the text). The lightest orange curve corresponds to $q_{\Gamma M}=0.2~$\AA$^{-1}$, while the darkest one to the $M$ point at $q_{\Gamma M}=0.97~$\AA$^{-1}$. Left and right panels correspond respectively to the full and zoomed spectra.}
    \label{fig:eels_d}
\end{figure*}

As mentioned in the main text, we use a small binning to reduce the noise in the recorded EELS measurements. Still, a smaller level of fluctuations remains, which we remove by applying a Savitzky-Golay filter with a polynomial of 4 order and a windows size of 31. The results of this filter for the momentum-resolved data along a subinterval of the $\Gamma M$ direction are plotted in Fig.~\ref{fig:eels_d}. In the left panel we show the full spectra before and after the filter, while in the right panel we show a zoom in a specific region where the noise is visible. 

\begin{figure*}
    \centering    
    \includegraphics[width=0.98\textwidth]{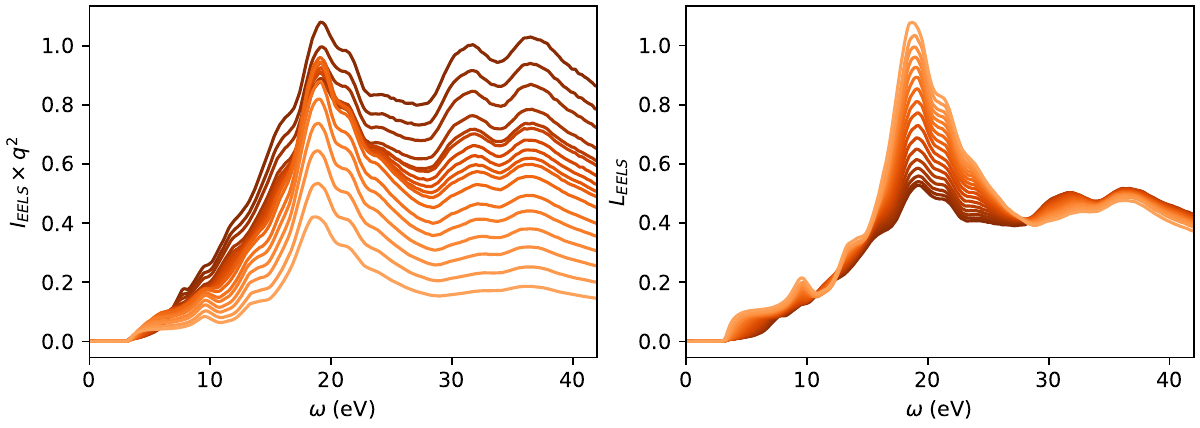}
    \caption{(left) Denoised spectra from Fig.~\ref{fig:eels_d} rescaled by a momentum-dependent factor $q^2$. (right) Loss function spectra obtained by a momentum-dependent normalization based on the f-sum rule (see details in the text). As in Fig.~\ref{fig:eels_d}, the curves in orange shades in both panels correspond to a subinterval of the $\Gamma M$ direction, from $q_{\Gamma M}=0.2~$\AA$^{-1}$ (lightest) to $q_{\Gamma M}=0.97~$\AA$^{-1}$ (darkest).}
    \label{fig:eels_n}
\end{figure*}

As also shown in the right panel of Fig.~\ref{fig:eels_d}, the measured intensity has an arbitrary scale depending on the experimental setting, which decreases rapidly with increasing momentum, $\mathbf{q}$. This dispersion corresponds to the double differential cross section defined in Eq.~(1) of the main paper, which is proportional to the product of the loss function, $L(\mathbf{q},\omega)$, and the Coulomb potential, $v(\mathbf{q})$. However, the loss function cannot be straightforwardly factorize from this relation since the $\mathbf{q}$ dispersion is affected by broadening due to e.g. finite energy and momentum resolutions. In Fig.~\ref{fig:eels_n} (left panel) we shown the results of normalizing the measured intensity by a quadratic factor, $q^2$, given by the Coulomb potential. The obtained quantity has a different dispersion with respect to Fig.~\ref{fig:eels_d}, for example, the intensity of the plasmon increases with $q$, and therefore it cannot be identified as the loss function. As shown in Fig.~\ref{fig:eels_n} (left panel), we obtain a qualitatively correct dispersion with the procedure described in App.~A of the main paper to extract the loss function from the experimental data, based on the f-sum rule.

\bibliography{biblio}